\documentclass[useAMS,usegraphicx,usenatbib]{mn2e}

\usepackage{amssymb}
\usepackage{times}
\usepackage{epsfig}

\usepackage{longtable}

\newcommand{\be}{\begin{equation}}
\newcommand{\ee}{\end{equation}}
\newcommand{\beq}{\begin{eqnarray}}
\newcommand{\eeq}{\end{eqnarray}}

\def\Rbb{R_{\rm bb}}
\def\Rns{R}
\def\Rapp{R_{\rm bb}} % the same as Rbb!!
\def\fc{f_{\rm c}}
\def\Teff{T_{\rm eff}}
\def\Tc{T_{\rm c}}

\def\Mdot{\dot{M}}
\def\Ledd{L_{\rm Edd}}

\def\Fper{F_{\rm per}}
\def\Ftdmean{\langle F_{\rm td} \rangle}
\def\Ftd{F_{\rm td}}
\def\Fedd{F_{\rm Edd}}

\def\Tbb{T_{\rm bb}}

\def\NH{N_{\rm H}}
\def\ergcm2s{{\rm erg\,cm^{-2}\,s^{-1}}}
\def \xte {{\it RXTE}}

\def\araa{ARA\&A}%
\def\apj{ApJ}%
\def\apjl{ApJ}%
\def\aap{A\&A}%
\def\mnras{MNRAS}%
\def\pasj{PASJ}%
\def\apjs{ApJS}%
\def\aapr{A\&ARv}		  
\def\ssr{Space Sci. Rev.}		  
\def\physrep{Physics Reports}		  
\def\prd{Physical Review D}		  
\title[The influence of accretion geometry on the spectral evolution during
thermonuclear (type-I) X-ray bursts]{The influence of accretion geometry on the spectral evolution during
thermonuclear (type-I) X-ray bursts}
\author[Jari J. E. Kajava et al.]
{Jari\,J.\,E.\,Kajava,$^{1,2,3}$\thanks{E-mail: jkajava@sciops.esa.int}
Joonas N\"{a}ttil\"{a},$^{3,4}$ Outi-Marja Latvala,$^{3}$ Miika
Pursiainen,$^{3}$ 
\newauthor
Juri Poutanen,$^{3,4}$ Valery F. Suleimanov,$^{5,6}$ Mikhail G.
Revnivtsev,$^{7}$ \newauthor Erik Kuulkers$^{1}$ and Duncan K.
Galloway$^{8,9,10}$\\
$^1$European Space Astronomy Centre (ESA/ESAC), Science Operations
Department, 28691 Villanueva de la Ca\~{n}ada, Madrid, Spain\\
$^2$Nordic Optical Telescope, Apartado 474, 38700 Santa Cruz de La Palma, Spain
\\
$^3$Astronomy Division, Department of Physics, P.O.Box 3000, 90014 University of
Oulu, Finland \\
$^4$Tuorla Observatory, University of Turku, V\"{a}is\"{a}l\"{a}ntie 20, 
FIN-21500 Piikki\"{o}, Finland\\
$^5$Institut f\"{u}r Astronomie und Astrophysik, Kepler Center for Astro and
Particle Physics, Universit\"{a}t T\"{u}bingen, Sand 1, D-72076 T\"{u}bingen,
Germany\\
$^6$Kazan (Volga region) Federal University, Kremlevskaja str., 18, Kazan
420008, Russia\\
$^7$Space Research Institute, Russian Academy of Sciences, Profsoyuznaya 84/32,
117997 Moscow, Russia\\
$^8$Monash Centre for Astrophysics, Monash University, Clayton, Victoria 3800,
Australia\\
$^{9}$School of Mathematical Sciences, Monash University, Clayton, Victoria
3800,
Australia\\
$^{10}$School of Physics, Monash University, Clayton, Victoria 3800, Australia\\
}

\begin{document}

%\date{Accepted. Received; in original form}

\pagerange{\pageref{firstpage}--\pageref{lastpage}} \pubyear{2014}

\maketitle

\label{firstpage}

\begin{abstract}
Neutron star (NS) masses and radii can be estimated from observations of
photospheric radius-expansion X-ray bursts, provided the chemical composition of
the photosphere, the spectral colour-correction factors in the observed
luminosity range, and the emission area during the bursts are known. By
analysing 246 X-ray bursts observed by the Rossi X-ray Timing Explorer from 11
low-mass X-ray binaries, we find a dependence between the persistent spectral
properties and the time evolution of the black body normalisation during the
bursts. All NS atmosphere models predict that the colour-correction factor
decreases in the early cooling phase when the luminosity first drops below the
limiting Eddington value, leading to a characteristic pattern of variability in
the measured blackbody normalisation. However, the model predictions agree with
the observations for most bursts occurring in hard, low-luminosity, ‘island’
spectral states, but rarely during soft, high-luminosity, ‘banana’ states. The
observed behaviour may be attributed to the accretion flow, which influences
cooling of the NS preferentially during the soft state bursts. This result
implies that only the bursts occurring in the hard, low-luminosity spectral
states can be reliably used for NS mass and radius determination.
\end{abstract}


\section{Introduction}\label{intro}

\begin{figure*}
\begin{center}
\epsfig{file=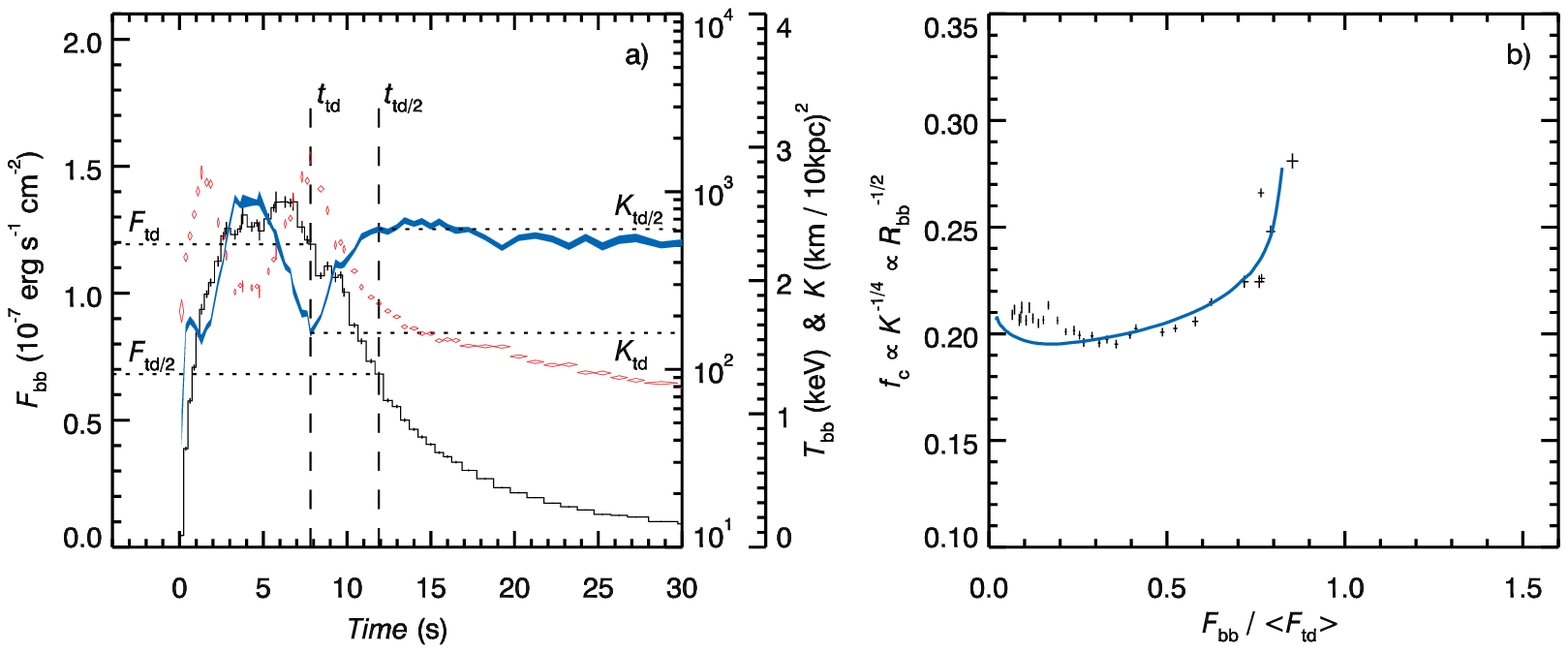}
\epsfig{file=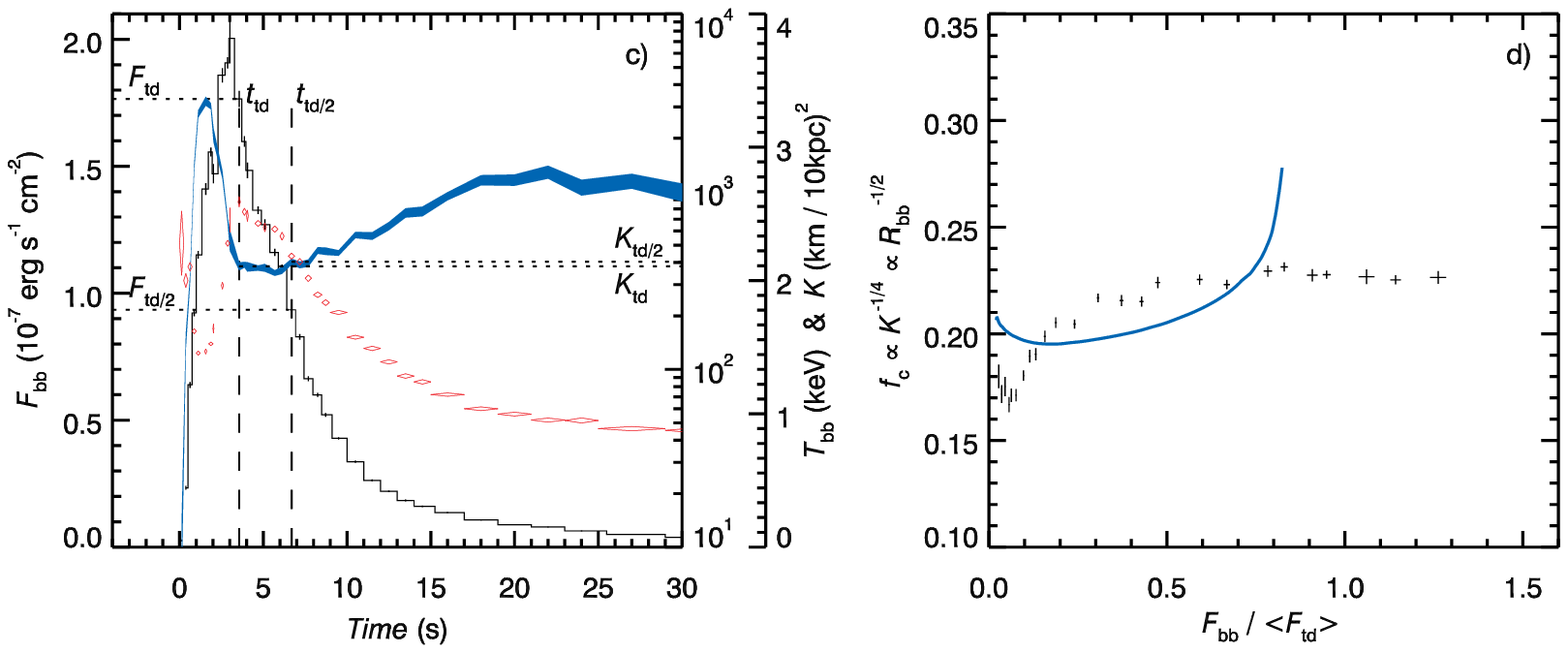}
\end{center}
\caption{Time resolved spectroscopy of two PRE X-ray bursts from 4U 1608--52
illustrating the differences between hard- and soft state X-ray bursts. 
In panels a) and c), the black line shows the bolometric flux $F_{\rm bb}$ in
units of $10^{-7}\,\ergcm2s$ (left-hand y-axis).
The blue ribbon shows the $1\sigma$ limits of the black body normalisation 
$K = (R_{\rm bb} [{\rm km}] / d_{10})^2$ (inner right-hand y-axis). 
The red diamonds show the $1\sigma$ errors for black body temperature $\Tbb$ in
keV (outer right-hand y-axis). 
The first black vertical dashed line marks the time of touchdown $t_{\rm td}$
and the second vertical dashed line to the right shows the time
$t_{\rm td/2}$ when $F_{\rm bb}$ has decreased to one half of the touchdown
flux.
The corresponding $F_{\rm bb}$ and $K$-values at these times $\Ftd$,
$F_{\rm td/2}$, $K_{\rm td}$ and $K_{\rm td/2}$ are marked with dotted lines.
The panels b) and d) show the relationship between the inverse square root of
the black body radius (proportional to the colour-correction factor $\fc$) and
the black body flux $F_{\rm bb}$ that is scaled using the mean touchdown flux
$\Ftdmean$.
The blue line is a model prediction for a pure hydrogen NS atmosphere with a
surface gravity of $\log g = 14.3$, taken from \citet{SPW12}.
The atmosphere model is the same for both b) and d) panels and it is shown
here to illustrate how well (or poorly) it follows the observed data.
Other NS atmosphere models computed for different chemical compositions 
and $\log g$ values can describe the hard state burst as well 
(see \citealt{SPW12}, figures 8 and 9).
Note that for this particular source $\Ftd$ is strongly variable between
bursts making the determination of $\Fedd$ non-unique.
Note also that because of telemetry issues, there are gaps in the high time
resolution data around $\Ftd$ that sometimes make touchdown time $t_{\rm
td}$ difficult to determine.}
\label{fig:burst}
\end{figure*}

More than 100 low-mass X-ray binary systems (LMXB) are known to produce
thermonuclear (type I) X-ray bursts (see \citealt{GMH08}).\footnote{See also the
burster list in
www.sron.nl/~jeanz/bursterlist.html‎}
These X-ray bursts are powered by unstable nuclear burning of helium and/or
hydrogen into heavier elements in the neutron star (NS) `ocean' (see
\citealt{LvPT93,SB06}, for a review).
Observationally they appear as rapid X-ray flashes, where the X-ray emission
increases by a factor of up to $\sim\!100$ in a few seconds, depending on the
persistent emission level.
After reaching the peak, the X-ray flux typically decays back to the persistent
level within a few tens of seconds. 
A fraction of X-ray bursts are so energetic that the Eddington limit is
reached, which causes the entire NS photosphere to expand.
This expansion can be seen through time resolved X-ray spectroscopy of the X-ray
burst (see e.g., Figure \ref{fig:burst}).
Close to the peak flux the photospheric radius expansion (PRE) causes
a characteristic decrease of the observed black body colour temperature $\Tbb$
with a simultaneous increase in the emission area \citep{HCL80,GMH80}.
These PRE-bursts are a very interesting sub-class of X-ray bursts, because they
can be used to constrain NS masses and radii (see e.g.,
\citealt{FT86,LvPT93,OGP09,SPR11}) and to estimate distances to the bursters
(e.g., \citealt{KdHitZ03} and references therein).
Accurate mass and radius measurements provide a way to probe the properties of
ultra-dense matter found in the cores of NSs through a careful comparison with
NS model predictions.
Thus, these measurements can be used to determine whether exotic particles, such
as hyperons, pion- or kaon condensates, or even de-confined quark matter form
when nucleons are compressed beyond their equilibrium nuclear densities
\citep{LP07,HPY07}. 

Most of the NS mass and radius measurements rely on the observation that the
X-ray emission comes from the NS surface and that the spectrum can be adequately
fitted with a simple thermal black body model (e.g., \citealt{GMH08,GPO12}). 
This way the observed colour temperature and flux can be associated with a
spherical black body radius $\Rbb$ \citep{LvPT93}.
Assuming the entire NS surface is burning during the X-ray burst -- and that
the entire NS is visible (rather than being obscured by the accretion disc) --
then the black body radius is related to the NS radius $\Rns$ through (see
\citealt{LvPT93})
\be \label{eq:RNS}
\Rbb = \Rns (1 + z) \fc^{-2}, 
\ee
where $z$ is the gravitational redshift and $\fc$ is the colour-correction
factor, defined as the ratio of colour temperature $\Tc$ and the
effective temperature $\Teff$ of the photosphere ($\fc \equiv \Tc/\Teff$;
\citealt{LTH86}, \citealt{MJR04}, \citealt{MMJ05}, \citealt{SPW11},
\citealt{SPW12}).

The most commonly used technique to measure NS masses and radii is the so called
`touchdown method' (e.g., \citealt{DML90}).
It relies on two key assumptions. 
The first assumption is that at the touchdown point -- which is defined as the 
time $t_{\rm td}$ when the black body radius has a local minimum and the colour
temperature has a local maximum (see Figure \ref{fig:burst}) -- the luminosity
is at the Eddington limit, $\Ledd$, such that 
\be \label{eq:Ledd}
L_{\rm td} = \Ledd \equiv \frac{4\pi GMc}{\kappa} (1 + z)^{-1},
\ee
where $G$ is the gravitational constant, $M$ is the NS mass, $c$ is the light
speed and $\kappa$ is the opacity.
The second assumption is that the emission area is constant throughout the
cooling and that the photospheric colour-correction factor $\fc$ asymptotically
reaches a constant value of $\fc \approx 1.4$ (see e.g., \citealt{OGP09}),
which is taken from NS atmosphere model predictions by \citet{MJR04} and
\citet{MMJ05}.

However, recent NS atmosphere model calculations by \citet{SPW12} indicate that
the both of these assumptions are incorrect. 
The Eddington luminosity can be up to 10 per cent larger as indicated by
Equation (\ref{eq:Ledd}), because of the Klein-Nishina reduction of electron
scattering cross-sections from the commonly assumed Thomson value.
Also, the validity of the latter assumption can be robustly checked using the
existing observational data.
All NS atmosphere models predict that the observer should see changes in 
the black body radii during the initial cooling phases of (PRE) X-ray bursts.
This prediction arises from the fact that $\fc$ should decrease when the
atmosphere starts cooling and the emitted
luminosity drops from the Eddington value \citep{MJR04, MMJ05, SPW11,SPW12}.
The most recent models of \citet{SPW12} predict that $\fc$ is in the range
$1.8$--$1.9$ when luminosity is close to the Eddington luminosity $\Ledd$, and
then $\fc$ decreases to a range $1.4$--$1.5$ by the time the luminosity
has dropped to $0.5\Ledd$.
Because the black body radius is proportional to the colour correction factor
as $\Rbb \propto \fc^{-2}$, we can test if X-ray burst data indeed
follow the model predictions by studying the time evolution of $\Rbb$ during the
initial first few seconds of the cooling phase. 

The observed $\Rbb$ values are known to show variations during the cooling
phases.
Already in 1986, \citet{GHP86} showed that $\Rbb$ could vary by a factor of
$\sim\!2$ in X-ray bursts of EXO 0748--676.
Similar variations were found in early observations of 4U 1636--536 by
\citet{DJP89}, for 4U 1705-44 by \citet{GHL89} and for 4U 1608--52 by
\citet{NDI89}.
From early on, the observed $\Rbb$ variations were attributed to either
anisotropic emission, varying emission area, changes in the
colour correction factor or a varying photospheric chemical composition
\citep{LS85,SFvP87,DJP89,LvPT93}.
As the $\Rbb$ variations are correlated with cooling time scales, they likely
reflect a changing chemical composition of the NS atmosphere from burst to
burst \citep{DJP89,BMG10}, which can ultimately be caused by variations of
the mass accretion rate $\Mdot$ onto the NS surface (see \citealt{FHM81}).
Although advances have been made in recent years, it is still not clear which of
these processes drive the observed $\Rbb$ evolution
\citep{BMG10,ZMA11,GL12,GPO12}.

\begin{figure}
\begin{center}
\begin{tabular}{c}
\epsfig{file=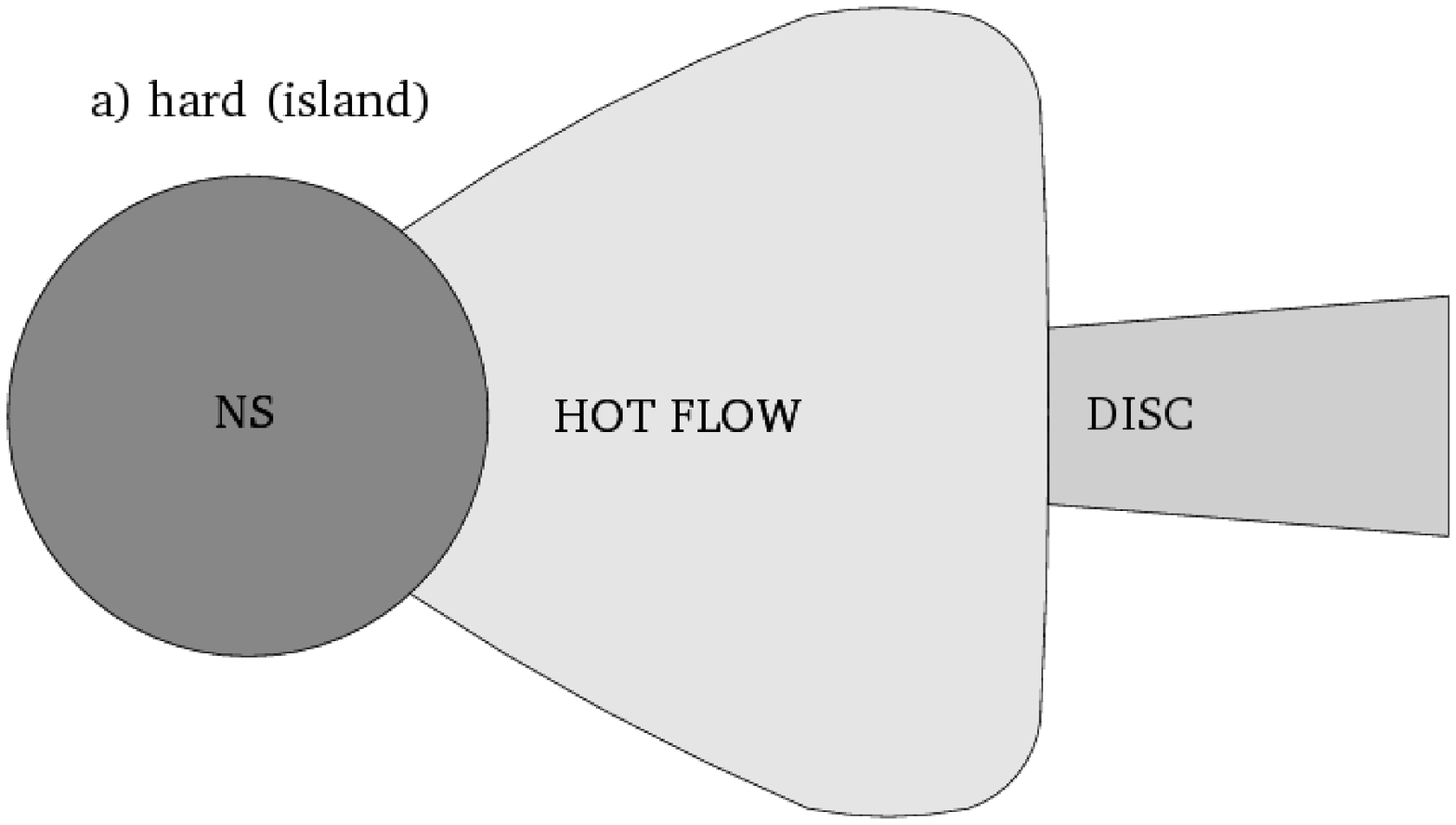,width=\linewidth} \\
\epsfig{file=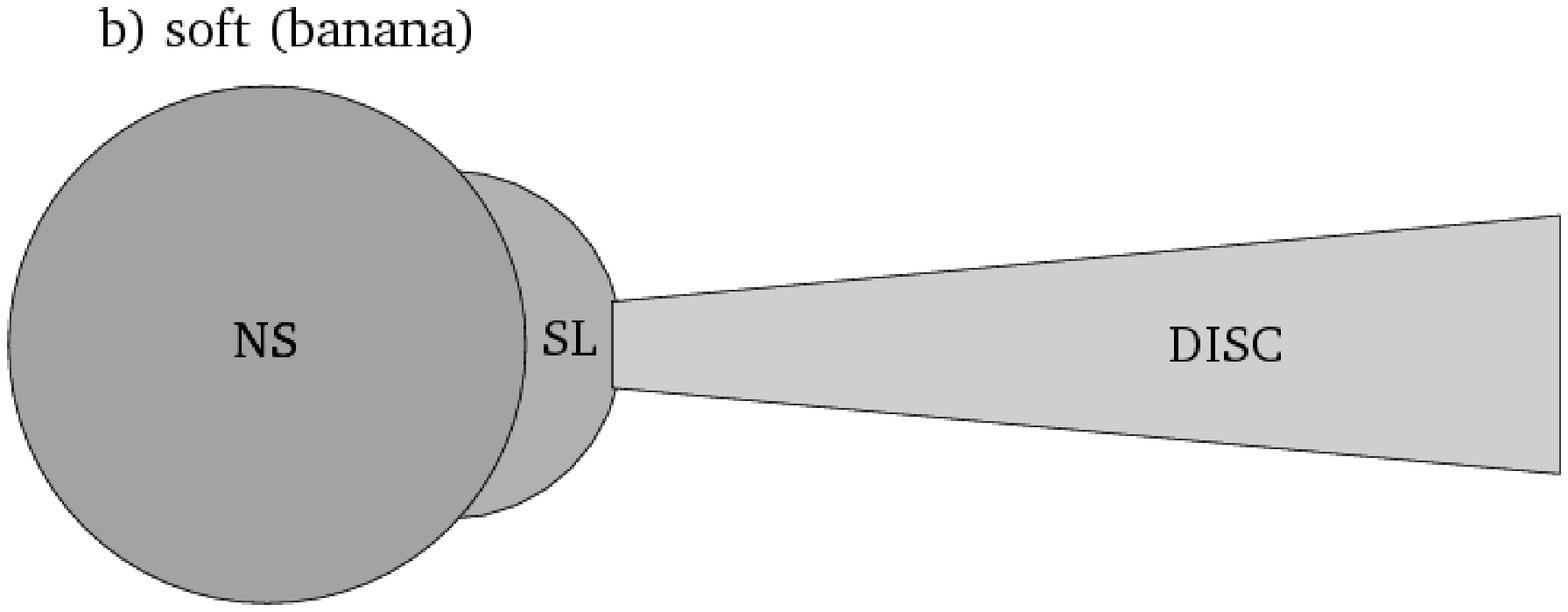,width=\linewidth} \\
\end{tabular}
\end{center}
\caption{An illustration of the assumed accretion geometries for the LMXBs that
do not show persistent pulsations. 
a) In the hard, island state the optically thick and geometrically thin
accretion disc is assumed to be truncated at some radius, and the innermost
accretion disc is assumed to form a hot, optically thin quasi-spherical flow.
b) In the soft, banana state the accretion disc is assumed to reach the NS
surface, where a spreading layer (SL) forms that can cover a fraction, or the
entire
the NS depending on the accretion rate \citep{IS99, SP06, IS10}.
Note that in the soft state the accretion disc and the spreading layer also
block the `other half' of the NS from view, 
whereas in the hard state this may happen only if the truncation radius is near
the NS surface
and the inclination angle is large.}
\label{fig:geometry}
\end{figure}

In this paper, we present a new X-ray burst diagnostic to test if and when the
NS cooling actually follows the atmosphere model predictions during X-ray
bursts.
In \S \ref{sec:observations} we first describe the sample of LMXBs we
have studied, and how we treated and modelled the observational data.
Then in \S \ref{sec:states} we show that the NS cooling is only
consistent with the models when the X-ray bursts take place during the
hard, `island' spectral state of the LMXBs.
Because different spectral states in LMXBs are thought to be caused by
variations
in geometry of the inner accretion disc (see Figure \ref{fig:geometry}, and
\citealt{DGK07} for a review), we then show in \S \ref{sec:flow} that
X-ray burst spectra are strongly influenced by the accretion flow in the soft, 
`banana' spectral state, where X-ray burst cooling behaviour rarely follow the
atmosphere model predictions.
We therefore propose that the observed $\Rbb$ trends and variations in
the initial cooling phases are driven by the spreading layer (SL; i.e. a
boundary layer) that engulfs the NS \citep{IS99,PS01,SP06,IS10}.
Thus, rather than observing a passively cooling NS atmosphere, we instead 
suggest that we are witnessing an actively accreting NS spreading 
layer that severely distorts the emitted X-ray burst spectra in the soft
spectral states.

\section{Observations and data reduction}\label{sec:observations}

We obtained all available {\it Rossi X-ray Timing Explorer} (\xte)/PCA
data from the HEASARC archive for each of the LMXBs listed in Table
\ref{tab:targets}.
The sample consists of two LMXB subclasses: atolls \citep{HvdK89} and accreting
millisecond pulsars (AMP; \citealt{PW12}).
We selected these sources, because they are known to exhibit bright X-ray bursts
at a large range of persistent flux levels \citep{GMH08}, thus enabling us to
investigate the effect of persistent accretion rate and spectral state
of the LMXB on the X-ray burst emission.
Furthermore, by comparing the bursting behaviour of the AMP SAX J1808.4--3658 
to the intermittent pulsars (Aql X--1 and HETE J1900.1--2455) and non-pulsating 
bursters, we can study if the magnetic field is important in shaping the 
cooling tracks of X-ray bursts.
We did not take bright Z-sources, like Cyg X-2 (e.g., \citealt{Smale98}) or GX
17+2 (e.g., \citealt{KHVdK02}), into the sample because the high accretion
rate -- and consequently the elevated persistent emission level -- in these
systems makes it difficult to confidently determine the shape of the burst
spectrum in the whole cooling track.

We identified the X-ray bursts using a similar method as in \citet{GMH08}.
Apart from finding the X-ray bursts already catalogued by them -- which
covered data from the beginning of the mission until June 3, 2007 -- 
we also detected several additional X-ray bursts observed after this date.
We also used similar criteria as \citet{GMH08} to check if the X-ray burst
showed signs of PRE (see \citealt{GMH08}, \S 2.3). 
If PRE was detected we included the X-ray burst to the analysis presented in
this paper.
However, during the analysis we had to exclude some bursts because of various
technical reasons.
For example, for 4U 1608--52 we excluded three bursts: one because $t_{\rm td}$
could not be determined due to telemetry gaps during the burst peak
(OBSID: 80406-01-04-08), one because PRE was only marginal (OBSID:
70059-01-08-00) and one anomalous, marginal PRE burst where the touchdown
occurs before the burst flux reaches the peak (OBSID: 94401-01-25-02).
Similarly for SAX J1808.4--3658 we had to exclude the majority of the bursts
that were affected by data gaps.
We also did not analyse X-ray bursts that were observed during spacecraft
slews, nor the few cases where the PCA data mode was such that the
determination of background and persistent emission spectra were not possible.

\begin{table} 
  \caption{The LMXBs studied in this work. 
  The bracketed numbers in column 3 denote the number of bursts
excluded from the sample. The mean touchdown flux values $\Ftdmean$ in
column 4 are given in units of $10^{-7}\,\ergcm2s$, see text. The known NS spin
frequencies $f_{\rm spin}$ are given in Hz (see \citealt{Watts12}, table 1
and references therein).}
 %\centering
  \label{tab:targets}
  \begin{tabular}{@{}lllll@{}}
  \hline
   \hline
Name 			& Alternative name  	& PRE bursts & $\Ftdmean$ &
$f_{\rm spin}$\\
\hline
4U 1608--52 		& 4U 1608--522 		& 18 (3) & 1.40 & 620\\
4U 1636--536 		& V801 Ara		& 81 (3) & 0.69 & 581\\
4U 1702--429 		& 4U 1702--42		& 5 (1)	 & 0.82 & 329\\
4U 1705--44 		& 			& 5 (3)	 & 0.39 & \\
4U 1724--307 		& 4U 1722-30		& 4	 & 0.58 & \\
4U 1728--34 		& GX 354--0		& 79 (15)& 0.95 & 363\\
4U 1735--44 		& V926 Sco		& 13	 & 0.38 & \\
4U 1820--30 		& 3A 1820--303		& 16	 & 0.59 & \\
\hline
Aql X--1	 	& V1333 Aquilae		& 12 (2) & 1.04 & 550
\\
HETE J1900.1--2455 	& 			& 10	 & 1.14 & 377\\
SAX J1808.4--3658	& 			& 3 (6)	 & 2.39 & 401\\
\hline
\end{tabular}
\end{table}

Altogether we analysed 246 PRE-bursts in our study (see Table
\ref{table:bigtable}).
The \xte/PCA data were reduced with the {\sc heasoft} package (version
6.12) and response matrices were generated using {\sc pcarsp} (11.7.1) task of
this package.
The time resolved spectra were extracted from the Event-mode data using initial
integration times of 0.25, 0.5, 1.0 or 2.0 seconds, depending on the peak count
rate of the burst ($>$6000, 6000--3000, 3000--1500, or $<$1500 counts per
second).
Then each time the count rate after the peak of the burst decreased by a factor
of $\sqrt{2}$, the time resolution was doubled to maintain approximately the
same signal-to-noise ratio.
A 16 second spectrum extracted prior to the start of the X-ray burst was
subtracted as the background for each burst (see \citealt{KHVdK02,GMH08}).
We note that in a recent work by \citet{WGP13} it was shown that
the burst spectra are better described statistically if the persistent
emission is allowed to vary during the bursts. Similar conclusions have been found in
numerous follow-up studies (see, e.g., \citealt{intZGM13, POB14, KBK14}).
However, accounting for this effect is not expected to give significantly different results 
because the persistent flux levels are less than 15 per cent of the peak fluxes 
for all the bursts we have analysed (see further discussion in Section 4.3).

We added standard 0.5 per cent systematic errors to the spectra
\citep{JMR06} and paid particular attention to correct for dead-time
effects by computing the effective exposure time for each time bin following the
approach recommended by the instrument
team.\footnote{http://heasarc.gsfc.nasa.gov/docs/xte/recipes/pca\_deadtime.html}
These spectra were fitted in {\sc xspec} \citep{Arnaud96} in a $2.5-25$ keV
range using Churazov-weighting \citep{CGF96} because some spectral channels had
very few counts.
However, we note that selecting this weighting was not particularly important as
we observed that results were affected only by $<\!1$ per cent compared to the
more commonly used $\sqrt{N}$ weighting (where $N$ is the number of counts in
the detector channel).
To describe the burst emission, we used a black-body model ({\sc bbodyrad})
multiplied by interstellar absorption ({\sc phabs} model in {\sc xspec}).
The best-fitting parameters were then the black body temperature
$T_{\mathrm{bb}}$ and the normalisation constant $K_{\mathrm{bb}} =
(R_{\mathrm{bb}} [\mathrm{km}] / d_{10}) ^2 $, where $d_{10} \equiv d / 10
\mathrm{kpc}$ is the distance in units of $10$ kpc. 
We initially let the absorption column density $\NH$ to vary, but
subsequently fixed it to the mean value over the burst.
In the end, the fits using the constant mean $\NH$ values were chosen
for further analysis in order to minimise the number of free parameters.
Different choices for the value of the absorption column densities were
also studied but no significant deviations from the mean $\NH$ value
method was found.
Errors of these parameters were taken as $1\sigma$ confidence levels, which were
obtained with the {\sc error} task by finding $\Delta \chi^2 = 1$.

We describe the spectrum of the persistent emission through X-ray colours,
which we determine using a similar approach as \citet{DG03}.
Instead of obtaining X-ray colours from X-ray count rates in specific \xte/PCA
detector channels (as is commonly done, see e.g., \citealt{GMH08}), we instead
computed them from model fluxes using {\sc xspec}.
For each \xte/PCA observation we divided the standard 2 data into segments
of 160 seconds. 
We found the best fitting model using a simple procedure, where we initially
started fitting the data with an absorbed {\sc powerlaw} model.
We continued to add and replace model components (such as {\sc gaussian},
{\sc bbodyrad}, {\sc diskbb} and {\sc comptt}), until finding a model that could
not be rejected with a higher than 95 per cent probability.
In these fits we always fixed the {\sc gaussian} line energy to $6.4$ keV, {\sc
comptt} optical depth to $\tau = 1$, the seed photon temperature to the {\sc
diskbb} temperature (when used together in the same model) and required the
hydrogen column density to be within an order of magnitude of the Galactic line
of sight column $N_{\rm H, gal}$, determined using the {\sc nh} ftool.
We also tried to let the {\sc comptt} seed photon temperature be a free
parameter as a last resort fit, but in the majority of cases where an acceptable
fit were not found before this, even this model did not provide a statistically
acceptable fit.
In these cases -- which were approximately 3.6 per cent of all PCA spectra
-- we chose to ignore these spectra altogether as the colour-colour diagrams
are only used for illustrative purposes.
After finding the simplest best fitting model (with the smallest number of free
parameters), we computed the unabsorbed fluxes using the {\sc cflux} model in
{\sc xspec} in four energy bands: $3$--$4$ keV, $4$--$6.4$ keV,
$6.4$--$9.7$ and $9.7$--$16$ keV. 
We then defined the hard and soft X-ray colours as flux ratios of $(9.7-16) /
(6.4-9.7)$ keV and $(4-6.4) / (3-4)$ keV, respectively.
We also computed the persistent flux $F_{\rm per}$ from the same best fitting
model over the $2.5\!-\!25$ keV band.
This method of computing the X-ray colours has the distinct advantage that it
helps to mitigate the effect of interstellar absorption that causes differences
the values of the soft colour in different LMXB systems. 
Also, this method eliminates small errors in X-ray colours due to PCA
gain change related variations in energy-to-channel conversions over the
lifetime of \xte.

\section{X-ray bursts during different spectral states}\label{sec:states}

\subsection{X-ray bursts and colour-colour diagrams}

Colour-colour diagrams (CC-diagrams hereafter) provide a model independent way
to describe spectral evolution of LMXBs.
Together with fast variability properties, the CC-diagrams provide the basis for
classifying LMXBs into atoll sources and Z-sources \citep{HvdK89}.
AMPs are typically put into the atoll category \citep{vSvdKW05}, but they are
different from normal atolls as they show persistent pulsations and their
CC-diagrams are also distinct. 
In our sample, we only have atoll sources as they tend to show X-ray bursts in
a large range of luminosities. 
The two intermittent AMPs -- HETE J1900.1--2455 \citep{GMK07} and Aql X--1
\citep{CAP08} -- are occasionally seen as AMPs, but otherwise behave like
normal atolls.

The CC-diagrams for the sources in our sample are shown in Figure
\ref{fig:ccdiagrams}.
The two distinct spectral states of atolls are easily visible in the
CC-diagrams.
The hard spectral state (top of the CC-diagram) is called the `island state',
whereas the soft state is called the `banana branch'
\citep{HvdK89}.
As in black hole transients, the hard states tend to be seen when luminosities
are low (hence at lower inferred mass accretion rates) and soft states are seen
at higher
luminosities (see right panels of Figure \ref{fig:ccdiagrams}, and
\citealt{DGK07} for a review).
For each source, the x-axis in the right hand panels of Figure
\ref{fig:ccdiagrams} denotes the persistent flux $\Fper$, which
is divided by the mean flux of all bursts of a given source at the touchdown
point $\Ftdmean$. 
As $\Ftdmean$ is close to the Eddington flux $\Fedd$, the
ratio is approximately $\Fper/\Ftdmean \approx \Fper / \Fedd$ and it can be
used as a proxy of the mass accretion rate $\Mdot$.\footnote{Note that this
ratio is similar to the commonly used $\Mdot$-proxy $\gamma \equiv
\Fper/ \langle F_{\rm peak} \rangle$, where $\langle F_{\rm
peak} \rangle$ is the mean peak flux \citep{vPPL88,GMH08}.}
However, we note that there are a few sources in our sample where $\Ftdmean$
shows large scatter (see Fig. \ref{fig:burst} and Table \ref{table:bigtable}),
so the estimate of $\Ftdmean$ can suffer from up to $\sim\!\!30$ per cent
systematic inaccuracy (see also \citealt{GMH08} for discussion).
In addition, $\Fper$ is also computed only in limited $2.5\!-\!25$ keV band.
In the soft states, when the emission can be often modelled with two black-body
like components, most of the flux is emitted in this band, but in hard
states $\Fper$ can underestimate the bolometric flux by up to a factor of two
\citep{GMH08}.
Because these large systematics dominate the error budget, we do not display
error bars and note that the uncertainty in the $\Fper/\Ftdmean$ values can be
up to factor of two in the hard state.

The X-ray colours that were observed right before the X-ray burst are
highlighted with coloured symbols. We used the bottom left area of the
CC-diagram as the dividing line below which bursts are referred as soft state
X-ray bursts. These bursts are displayed with red diagonal crosses. 
The hard state bursts at the top of the CC-diagram are denoted with green
crosses and the `intermediate' state bursts that lie in between these two well
defined states are shown with blue asterisks.
In a few sources, like 4U 1608--52, 4U 1636--536 and 4U 1728--34, the boundary
between the soft and the intermediate bursts is not straight forward to define.
In addition, for 4U 1728--34 the boundary between the hard and the intermediate
bursts is not well defined either.
In these sources the dividing lines were chosen somewhat arbitrarily, but the
results, or the conclusions are not affected if a few bursts are moved from one
category to the other.

\begin{figure*}
\begin{center}
\begin{tabular}{c}
\epsfig{file=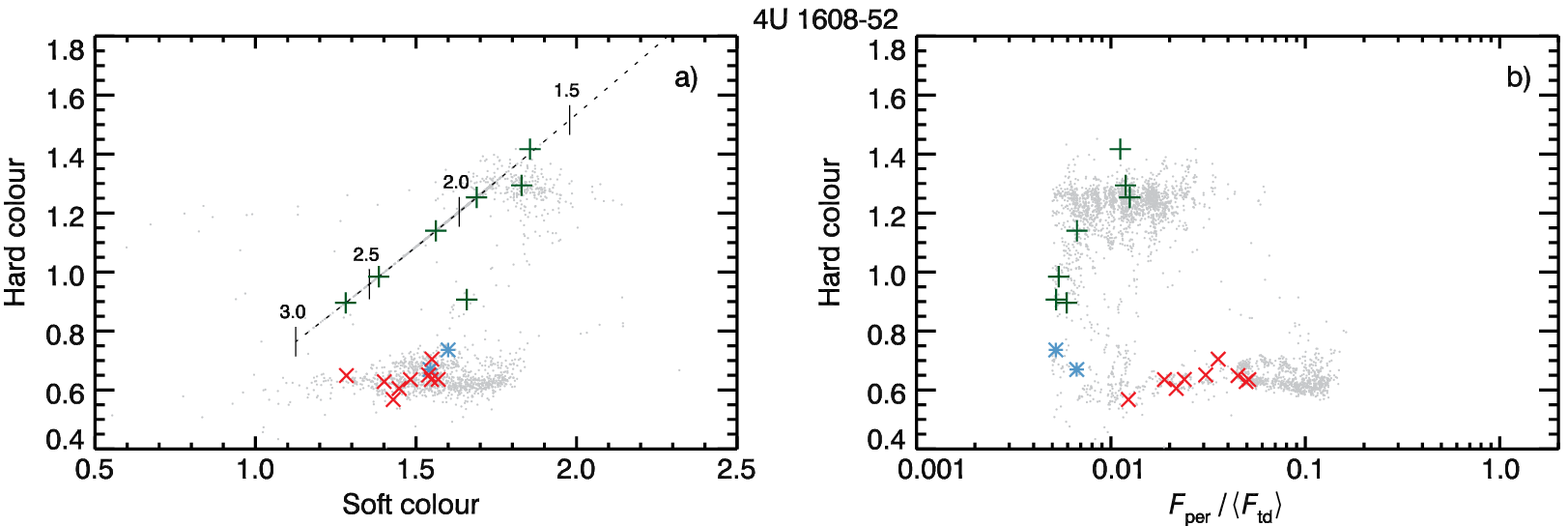} \\
\epsfig{file=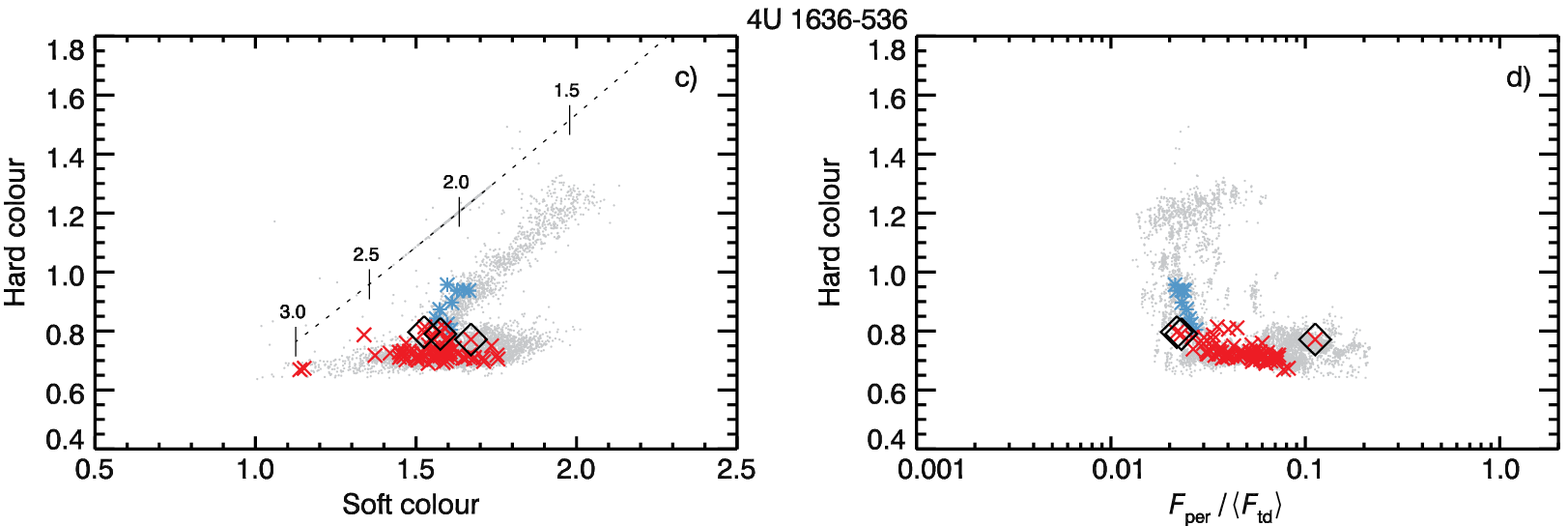} \\
\epsfig{file=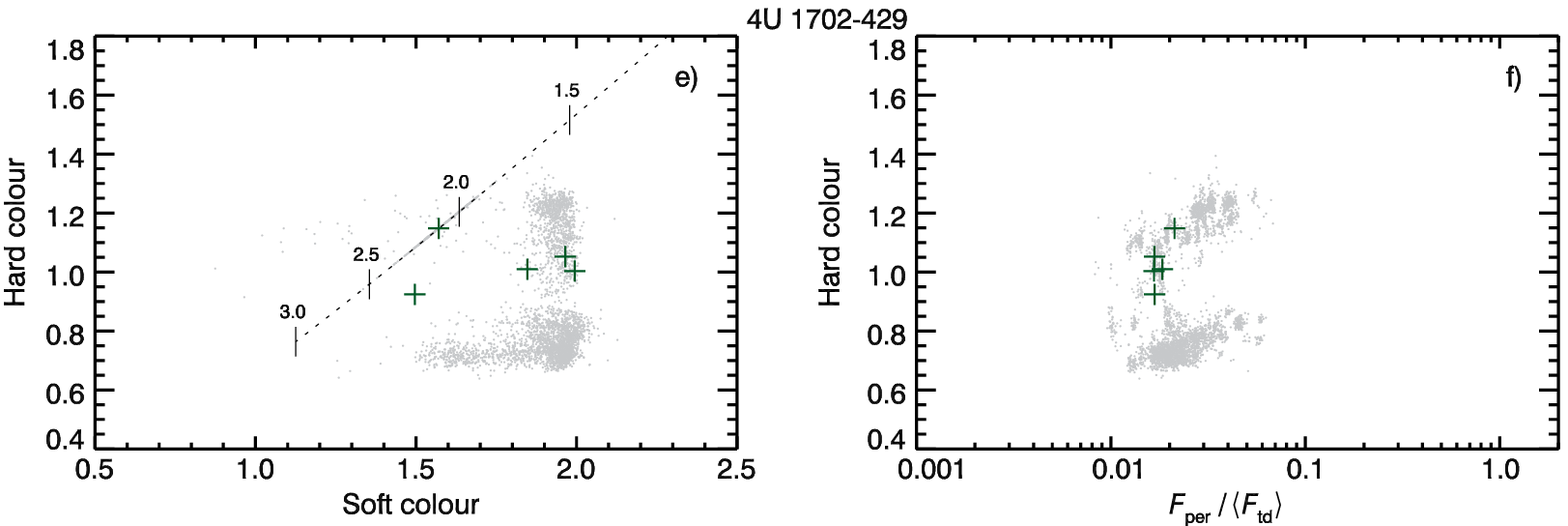} \\
\end{tabular}
\end{center}
\caption{Left panels: colour-colour (CC) diagrams for each LMXB in the studied
sample. The dotted line indicates the track for spectra consisting of an
absorbed power-law, with the spectral index $\Gamma$ indicated.
The hard- and the soft X-ray colours are flux ratios between $(9.7-16) /
(6.4-9.7)$ keV and $(4-6.4) / (3-4)$ keV, respectively.
The colours for the 160 second segments just before the PRE bursts are
highlighted with green crosses, red diagonal crosses and blue asterisks for PRE
bursts in hard, soft and `intermediate' states, respectively (the rest are
marked with grey dots). 
Right panels have the same y-axes as left panels, but the x-axes show the
persistent flux $F_{\rm per}$ scaled with the mean touchdown flux 
$\Ftdmean$.
Three bursts of 4U 1636--536 are highlighted using diagonal boxes for easier
cross-referencing with Figure \ref{fig:K2K1}.
}
\label{fig:ccdiagrams}
\end{figure*}

\begin{figure*}
\begin{center}
\begin{tabular}{c}
\epsfig{file=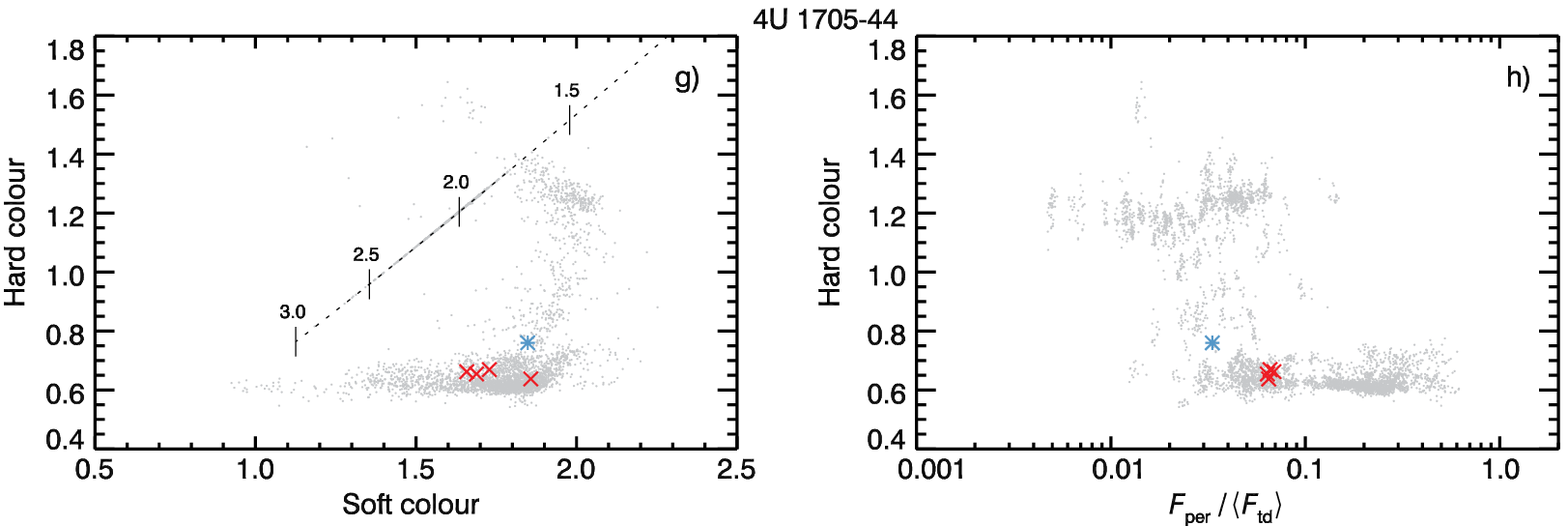} \\
\epsfig{file=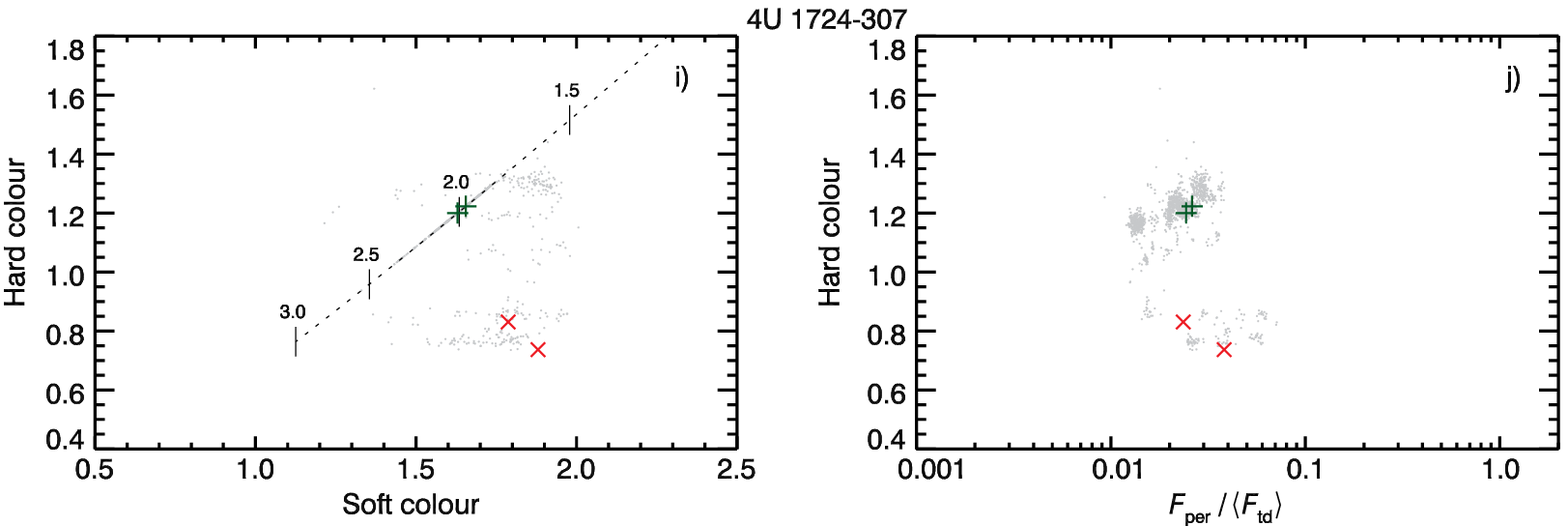} \\
\epsfig{file=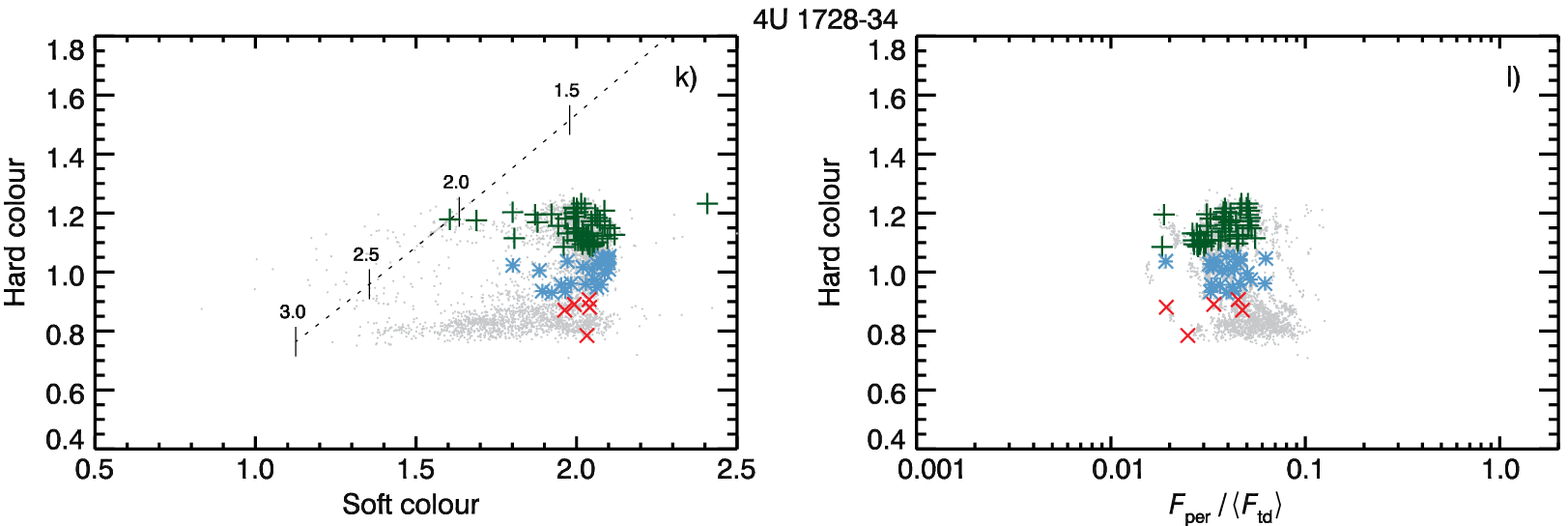} \\
\epsfig{file=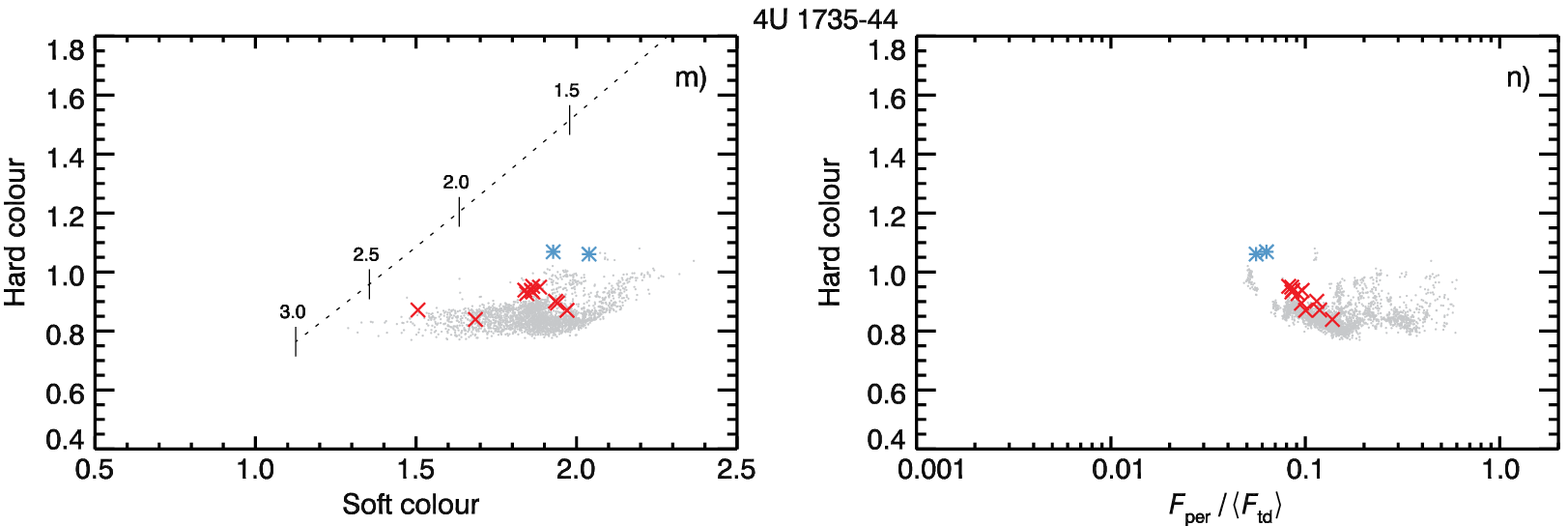}
\end{tabular}
\end{center}
\contcaption{}
\label{fig:ccdiagrams2}
\end{figure*}

\begin{figure*}
\begin{center}
\begin{tabular}{c}
\epsfig{file=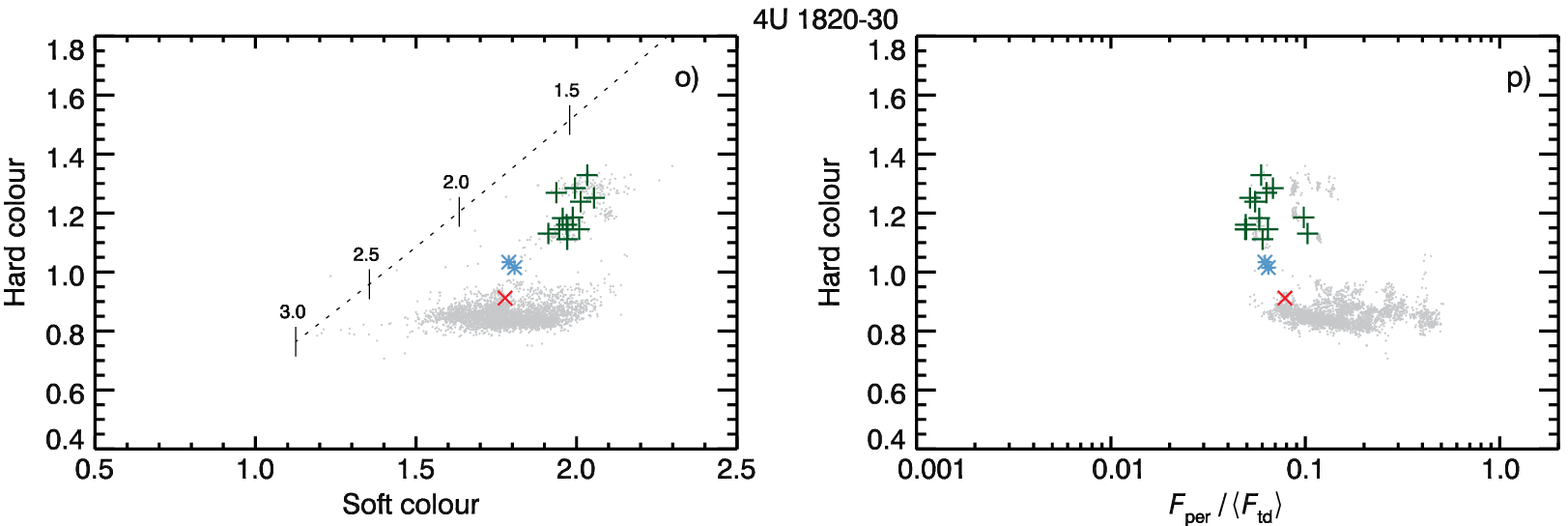} \\
\epsfig{file=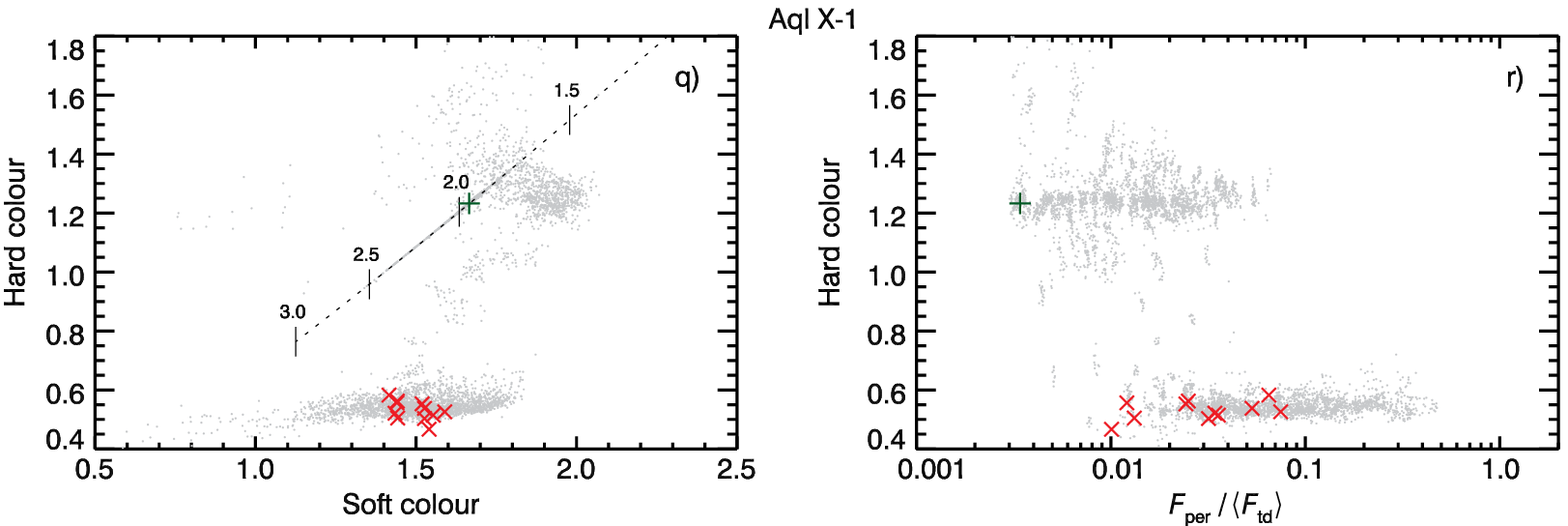} \\
\epsfig{file=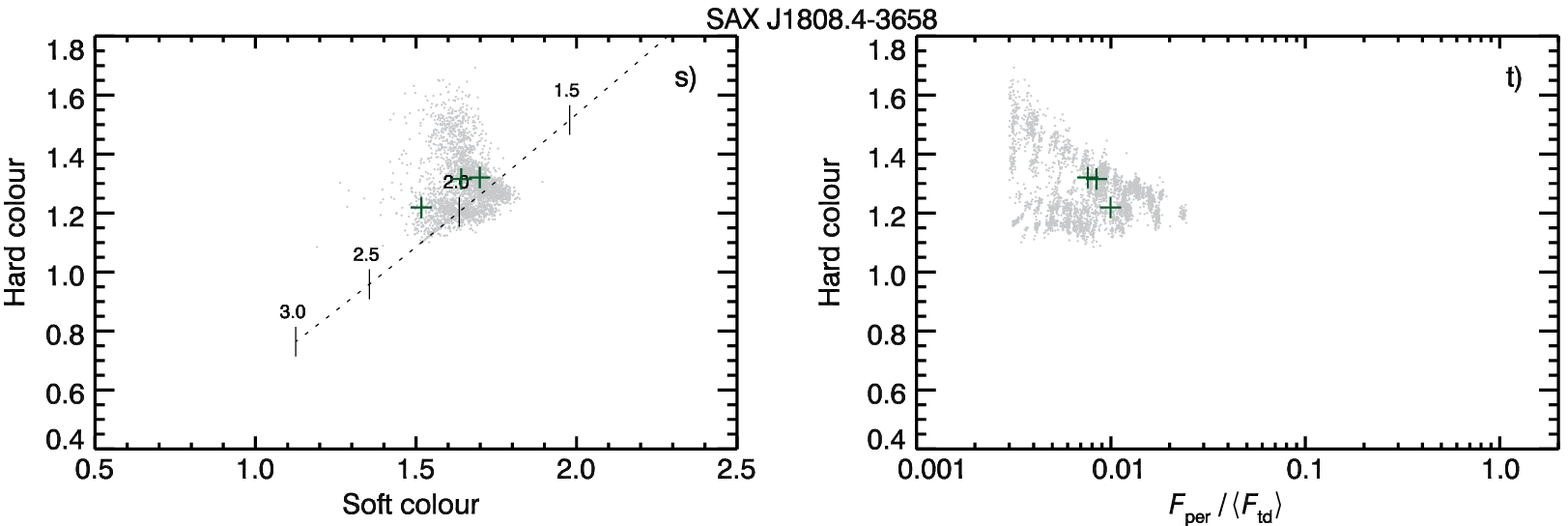} \\
\epsfig{file=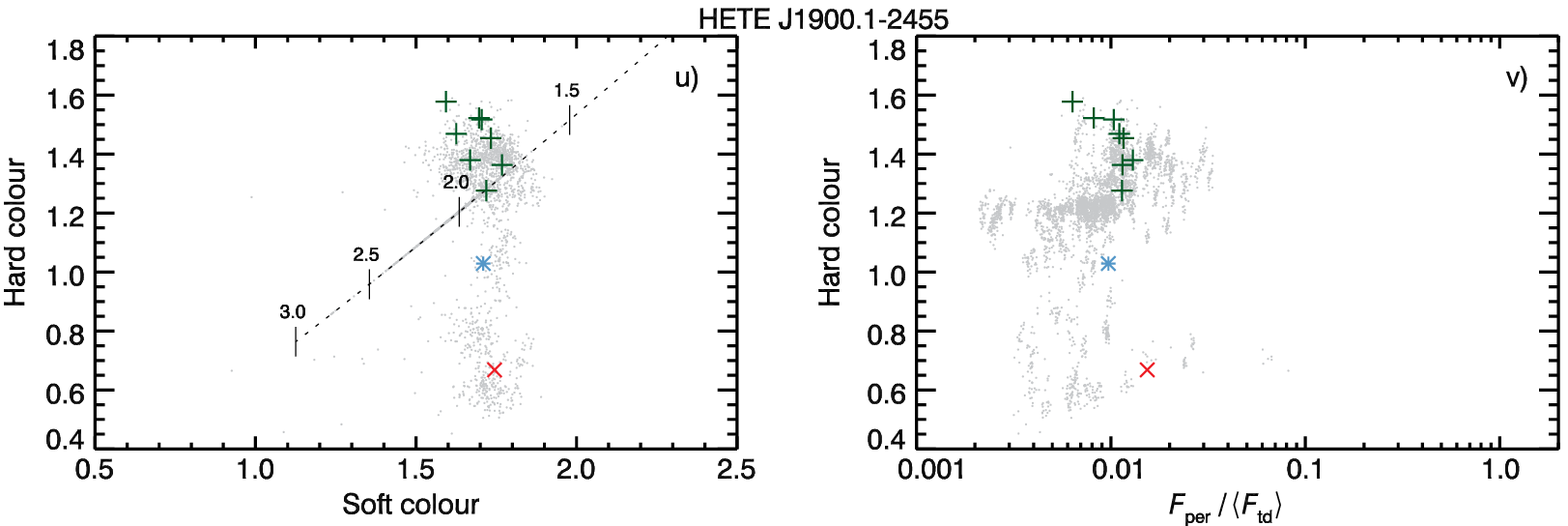}
\end{tabular}
\end{center}
\contcaption{}
\label{fig:ccdiagrams3}
\end{figure*}

One striking feature seen in Figure \ref{fig:ccdiagrams} is the diversity of
bursting behaviour among the sources in the sample. Different sources show
PRE-bursts in different regions of the CC-diagrams, and there are various
reasons causing the diversity. 
According to the standard burst theory \citep{FHM81,SB06}, steady nuclear
burning of hydrogen is thought to produce PRE-bursts in a pure helium layer in
the range of $\Fper/\Ftdmean \sim 0.01 - 0.05$, whereas for bursts above
$\Fper/\Ftdmean \gtrsim 0.05$ fresh accreted hydrogen builds up faster than it
can burn to helium, resulting in mixed hydrogen/helium bursts. However, these
dividing lines can vary from source to source for several reasons. Because the
relative amount of hydrogen and helium influences the burst energetics, one of
the main factors causing the diversity of bursting properties is the difference
in chemical composition of the accreted fuel between sources. 
There are two ultra-compact binaries which accrete hydrogen poor gas: 4U
1820--30 (a white dwarf--neutron star binary with an 11-minute period,
\citealt{SPW87}) and 4U 1728--34 (a candidate ultra-compact binary with a
possible $\sim\!\!10$ minute orbital period, \citealt{GYM10}), which sets them
apart from the rest of the sources in the sample.
Also, as noted by \citet{MGC04}, another key effect seems to be the NS spin
frequency: the `slowly' spinning systems such as 4U 1702--429 and 4U 1728--34
show PRE burst only (or preferentially) in the hard states, whereas faster
spinning sources such as 4U 1636--536 and Aql X--1 show PRE bursts
preferentially in the soft states (see also \citealt{GMH08}). The role of
turbulent mixing in the burning layers might be important to explain these
differences. According to \citet{PB07}, the mixing -- which is more effective at
higher accretion rates and at smaller NS spin frequencies -- can cause helium to
burn in steady state, and thus cause the bursting to stop in different parts of
the diagrams. These factors might be related to each other, because all the
slow-spinning sources for which the composition is known are hydrogen-poor,
while all the fast-spinning sources are hydrogen-rich \citep{GMH08}.
However, a detailed comparison is challenging because we do not know the spin-
nor orbital periods for all of the systems in the sample.
Furthermore, the sources make spectral state transitions in the same $\Mdot$
range where the bursting regimes are expected to change.
This may affect the bursting properties if the accretion geometry changes as
illustrated in Figure \ref{fig:geometry}.
The local mass accretion rate (i.e. $\Mdot$ per unit NS surface area), which is
important in defining the bursting properties, might change in an opposite way
during the transitions as indicated by the global $\Mdot$ (see, e.g.,
\citealt{Bildsten00}).
In addition, the presence of dynamically important magnetic fields in the AMP
SAX J1808.4--3658 -- and to some extent in the intermittent AMPs (Aql X--1 and
HETE J1900.1--2455) -- influences the accretion geometry, which further
complicates the comparison of the bursting behaviour between the sources in the
sample. 

Although the bursting behaviour is diverse and not necessarily fully
understood yet, we can nevertheless investigate how the evolution of the black
body normalisation (and thus $\Rbb$) depends on the spectral state the burst
occurred, given that a large number of PRE-bursts are detected in different
spectral states between the sources in the sample. 

\subsection{A new X-ray burst diagnostic: the K-ratio}\label{sec:diagnostic}

\begin{figure*}
\begin{center}
\epsfig{file=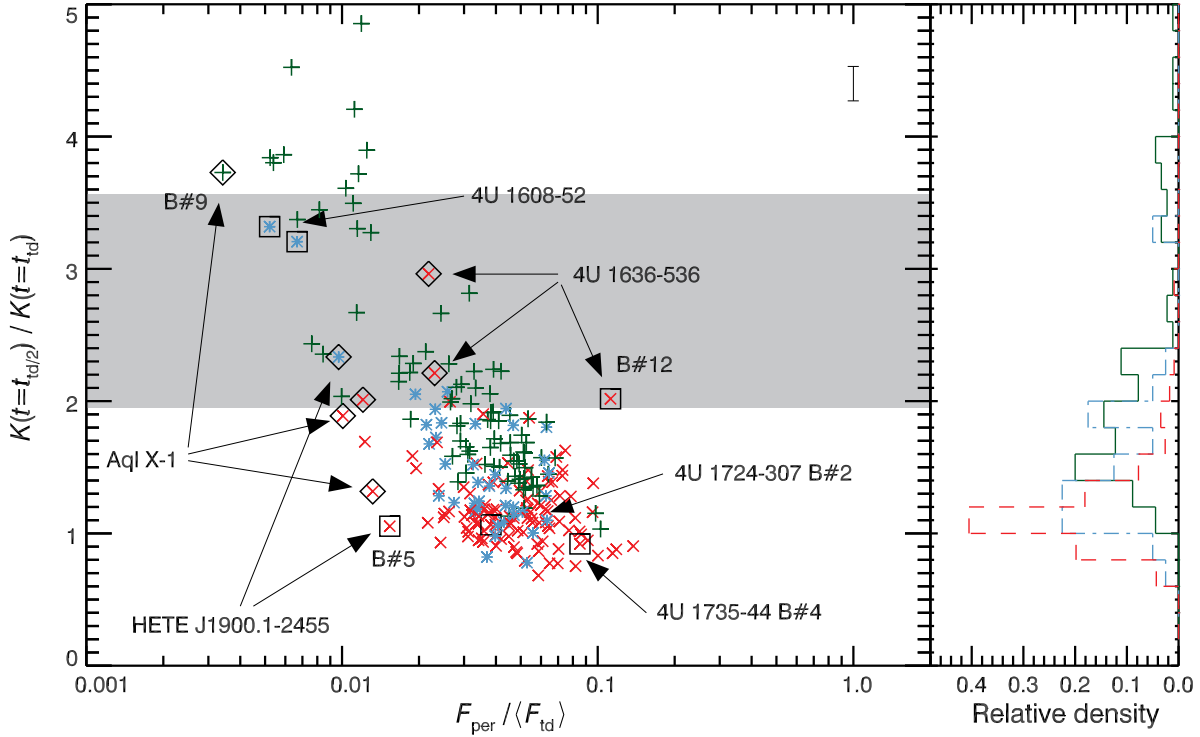}
\end{center}
\caption{Ratio of black body normalisations at `half-touchdown' and at
touchdown as a function of the scaled persistent flux 
$\Fper / \Ftdmean$.
The symbol colours are based on the CC-diagrams (left panels of Figure
\ref{fig:ccdiagrams}) with green crosses, red diagonal crosses and blue
asterisks denoting PRE bursts that occur during hard, soft and
`intermediate' states, respectively.
A typical $K$-ratio error bar (the mean of all $1\sigma$ $K$-ratio errors) is 
shown at the top right corner (see Table \ref{table:bigtable}).
Note also that the x-axis is the same as in the right hand side panels of Figure
\ref{fig:ccdiagrams} and that the way how $K$-values in the y-axis are computed
is illustrated in Figure \ref{fig:burst}.
The grey horizontal band denotes the area where $K_{\rm td/2} / K_{\rm
td} \propto (f_{\rm c}(t=t_{\rm td/2})/f_{\rm c}(t=t_{\rm td}))^{-4}$ ratio is
consistent with \citet{SPW12} models predictions (see text).
The right panel histograms show how bursts in different spectral states are
distributed differently on this diagram (hard, soft and
`intermediate' state histograms are marked with a green line, a red dashed
line and a blue dash-dotted line, respectively).
Only the hard state bursts that take place at persistent fluxes lower than
$\sim\!3$ per cent of the Eddington value are consistent with the model
predictions.
Those hard state bursts that have $K_{\rm td/2} / K_{\rm td} < 2$ are all from
two sources: 4U 1728--34 and 4U 1820--30.
A few interesting bursts that have significantly lower touchdown fluxes than
 $\Ftdmean$ are highlighted with squares, whereas other interesting `outlier'
bursts (with $\Ftd \gtrsim \Ftdmean$) are highlighted using diagonal boxes (see
discussion in Section 4.3).
}
\label{fig:K2K1}
\end{figure*}

Because of the large collecting area of \xte/PCA and the brightness of the
sources in the sample, we can perform time-resolved spectroscopy of the bursts
in the
initial cooling phases.
The most recent models of \citet{SPW12} predict that the colour correction 
factor $\fc$ should drop from $1.8$--$1.9$ at touchdown to $1.4$--$1.5$ before
the flux had dropped to half of it.
According to Equation (\ref{eq:RNS}) the black body normalisation $K \propto
\Rapp^2 \propto f_{\rm c}^{-4}$ and, therefore, to see when bursts follow
this expected behaviour, we devised a very simple diagnostic.
We extracted two black body flux and normalisation values from the data for
each X-ray burst (see Figure \ref{fig:burst}).
We first located the touchdown time $t_{\rm td}$, and the corresponding flux
and black body normalisation at touchdown $F_{\rm td} (t\!=\!t_{\rm td})$ and
$K_{\rm td} (t\!=\!t_{\rm td})$.
We then located the same values after the flux had dropped to half of the
touchdown flux, i.e. $F_{\rm td/2} (t\!=\!t_{\rm td/2})$ and $K_{\rm td/2}
(t\!=\!t_{\rm td/2})$, where $F_{\rm td/2}(t=t_{\rm td/2}) = 0.5
F_{\rm td}(t=t_{\rm td})$ (we took the values just before flux dropped
below $0.5 F_{\rm td}$).
We then took a ratio of these normalisation values $K_{\rm td/2} /
K_{\rm td} \propto (\Rapp(t\!=\!t_{\rm td/2}) / \Rapp(t\!=\!t_{\rm
td}))^2 \propto (f_{\rm c}(t\!=\!t_{\rm td/2})/f_{\rm c}(t\!=\!t_{\rm
td}))^{-4}$ because this ratio does not depend on the distance, nor the
unknown gravitational redshift.
It is important to note that our selection of using the $K$-value at 
``half-touchdown'' makes very little difference in the resulting ``$K$-ratios'':
Any $K$-value between $F \approx 0.4 - 0.6\, F_{\rm td}$ can be used, because both 
the observed $K$-values, and the $\fc$ values from atmosphere model predictions 
are relatively constant around $F_{\rm td/2}$ (see Figure \ref{fig:burst} and \citealt{SPW12}). 
We then plot the $K_{\rm td/2} / K_{\rm td}$ ratios of all the bursts in the
sample as a function of the scaled persistent flux $\Fper / \Ftdmean$ in 
Figure \ref{fig:K2K1}.
We use the same colour coding as in the CC-diagrams of Figure
\ref{fig:ccdiagrams}.
In Figure \ref{fig:K2K1} we also show the area where $K_{\rm td/2} / K_{\rm td}
\propto (f_{\rm c}(t=t_{\rm td/2})/f_{\rm c}(t=t_{\rm td}))^{-4}$ ratios are
consistent with \citet{SPW12} model predictions. 
Even if we allow the NS photosphere to change composition during the burst, the
$K$-ratios should be in a tight range between $\approx\!2.0 - 3.6$.

The PRE bursts that occur during hard- and soft states have clearly different
$K_{\rm td/2} / K_{\rm td}$ ratios.
We can rule out the possibility that the X-ray burst cooling tracks in
these two states come from the same underlying distribution, because
the chi-squared test on the hard- and soft state $K_{\rm td/2} / K_{\rm td}$
histograms gives $\chi^2 = 66.14$ for $19$ degrees of freedom (${\rm p-value}
\approx 4 \times 10^{-7}$; alternatively the value of Kolmogorov-Smirnov
D-statistic is $D=0.705$ and ${\rm p-value} < 10^{-8}$; see \citealt{NR07}). 
The majority of soft state bursts have $K_{\rm td/2} / K_{\rm td} \approx 1$,
and only 4 out of 116 of them have $K_{\rm td/2} / K_{\rm td} > 2$.

In contrast, about 40 per cent of hard state bursts have $K_{\rm td/2} / K_{\rm
td} > 2$ and all the hard state bursts that do not follow the model predictions,
i.e. that have $K_{\rm td/2} / K_{\rm td} < 2$, are from the two ultra-compact
sources: 4U 1728--34 and 4U 1820--30.
In these two cases at very close to Eddington flux there is a characteristic
kink upwards in the $F-K^{-1/4}$ diagrams, similar to the burst shown in
Figure \ref{fig:burst} b), but the behaviour later on is not consistent with
model predictions (see also \citealt{GZM13}, figure 6).
However, it is not clear why the hard state $K$-ratios in these hydrogen poor
systems are so different from the rest of the sample.
The results in Figure \ref{fig:K2K1} nevertheless show that only bursts that
occur during the hard state, and occur below $\sim\!3$ per cent of the
Eddington luminosity are consistent with the model predictions.

\section{Discussion} \label{sec:flow}

\subsection{Effects of the accretion flow}

If none of the X-ray bursts followed the NS atmospheric model predictions of
\citet{SPW12}, then we would have a strong reason to doubt that
some essential piece of physics is still missing from the models (like stellar
rotation). 
However, the results presented in Figure \ref{fig:K2K1} show that this is not
the case.
We can see that there are a large number of X-ray bursts that fall within
the band where they are consistent with the atmosphere models: most of them
being bursts that occur during low-luminosity, hard states.
Therefore one should rather be asking why, apart from a few outliers, the X-ray
bursts in the soft state do not follow the expected cooling behaviour.

The data clearly indicates that the emission area for the soft state bursts
remain approximately constant, that is, the $K$-ratio is $K_{\rm td/2} / K_{\rm
td} \! \approx \! 1$.
This can only be realised if the emitting area and the colour correction factor
stay constant during the initial cooling phase \citep{GL12}.
An interpretation of this behaviour was given in \citet{SPR11}, where
PRE-bursts of 4U 1724--307 were analysed.
This interpretation assumes that the two spectral states are caused by
changes in the accretion geometry, which is illustrated in Figure
\ref{fig:geometry}.
The origin of the constant black body radius could be related to the spreading
layer (SL, \citealt{IS99,IS10}), but we emphasise that the key point is not the
emission produced by the SL itself; it can be determined from the spectrum of
the persistent emission moments before the burst and it can be subtracted from
the burst emission (as is routinely done in X-ray burst analysis).
Rather, the SL changes the X-ray burst radiation spectrum, because the emission
from the X-ray burst must pass through this layer before reaching the observer.
As the accreted gas enters the SL with roughly Keplerian orbital frequency
(always larger than the NS spin frequency), the SL is supported both by
centrifugal forces and the radiation pressure force produced by energy
dissipation within the SL and by the X-ray burst. This makes the outermost
layers of the SL tenuous and, therefore, the resulting radiation spectrum has
colour correction factor of about $\fc \approx 1.6 - 1.8$ for a large range of
luminosities from $L\! \sim \! \Ledd$ to $L \! \sim \! 0.2 \Ledd$
\citep{SP06,RSP13}.
In our view, this can explain the irregular cooling behaviour of X-ray bursts in
the soft state where the black body radius $\Rbb$ is constant in the
aforementioned luminosity range (i.e. $K_{\rm td/2} / K_{\rm td}\! \approx \!
1$).
However, the constancy of $\Rbb$ in the soft state bursts does not mean that we
see the entire NS surface, as is commonly proposed (e.g., \citealt{LS14}).
The $\Rbb$ can also be constant in the soft state bursts if the SL always blocks
a constant fraction (about the half) of the NS from view.

Our results also show that the bursting behaviour of the AMP SAX J1808.4-3658 and 
the intermittent pulsars Aql X--1 and HETE J1900.1--2455 is similar to the 
non-pulsating bursters in the sample. All hard state bursts of these pulsars have 
$K$-ratios consistent with the atmosphere models. This may be 
attributed to two factors: 1) the burst properties are determined by the conditions 
deep in the NS ocean, which are not affected by the magnetic field even if it is important 
from the dynamical point of view outside of the NS and 2) the magnetic field 
is known to truncate the accretion disc to a large radius and, therefore, the whole NS 
surface is likely to be visible during the cooling track.
That is, even if the magnetic field channels the accretion flow to the stellar poles in AMPs,
its observational effect on the burst spectra does not seem important.
The bursts from HETE J1900.1-2455 are especially interesting in this respect. 
The hard state bursts have high $K$-ratios both when persistent pulsations were detected 
in 2005 (B\#1, see Table \ref{table:bigtable}), \footnote{We note that the cooling track of B\#1 
is unique. This burst will be analysed in detail in a future publication.} and after they disappered.
However, when the source entered the soft state in 2009, that burst (B\#5) has 
$K_{\rm td/2} / K_{\rm td}\! \approx \! 1$ similarly to the rest of the non-pulsating 
sources in sample. Furthermore, the only hard state burst of Aql X--1 has a high $K$-ratio 
whereas the soft state bursts have $K_{\rm td/2} / K_{\rm td}\! \approx \! 1$.
Therefore, the behaviour of these three sources indicates that the NS magnetic field 
is not as important in determining the shape of the cooling tracks of X-ray bursts as is 
the SL in the soft state.

In addition, we speculate that the SL causes also another observable feature in
the time evolution of $\Rbb$ values in the late cooling stages when $L \lesssim
0.5 \Ledd$ (see Figure \ref{fig:burst}d).
According to \citet{IS99,IS10}, the latitudinal width of the SL depends on the 
accretion rate and thus the luminosity.
However, when an X-ray burst occurs beneath the SL it provides additional
radiation support to this levitating layer.
When the NS atmosphere starts cooling immediately after the X-ray burst peak and
the luminosity starts decaying from the Eddington value, the SL above the NS
starts to be less and less supported by the radiation pressure produced by the
burst.
During the first moments after the X-ray burst, when the luminosity is close to
the Eddington value, the SL might cover the whole NS (see \citealt{IS99}, figure
7). 
Over the next few seconds, when the burst luminosity drops towards the level of
the persistent emission, the SL may instead cover a smaller fraction of the NS
surface.
The polar regions of the NS would then be gradually exposed to the observer as
the luminosity drops.
This polar region would have $\fc \! \approx \! 1.4$, whereas the equatorial
regions of the star would still be covered beneath the SL with $\fc
\approx 1.6 - 1.8$.
Such uncovering of the NS surface would have the net effect of gradually
decreasing the colour correction factor when the burst luminosity drops.
We speculate that this might cause the observed $\Rbb$ variations below
$\sim\! 0.2\,\Fedd$ in the burst presented in Figure \ref{fig:burst}(d) (see
also \citealt{Poutanen14}).
Similar behaviour is seen in the late stages of soft state bursts (and the
non-PRE hard state bursts) of 4U 1636--536 (see figure 7 in \citealt{ZMA11}).
Future theoretical research towards this direction would be of interest, because
if our interpretation is correct in a qualitative sense, then, in principle,
these soft state X-ray burst could be used to place constraints on how the
latitudinal width of spreading layers depend on the luminosity.

\subsection{Implications for NS mass and radius measurements}
\label{sec:implications}

% FT86, 4U 1636 shows bursts only in soft and intermediate states.
% Therefore, I lump them to soft state.

% vPL87 took their data from Haberl et al. 1987, from where one can
% deduce that the bursts occured at ~10% Eddington.

% SFvP87 bursts occur at ~6% eddington and powerlaw is a worse fit than
% thermal bremss. Therefore, I suppose soft state bursts.

The most profound implication of our result has to do with the NS mass and
radius constraints that can be derived from the PRE-bursts. 
Apart from rare exceptions (such as the analysis presented in \citealt{vPDT90}),
there are many examples in the literature where soft state X-ray bursts have
been used to make NS mass-radius estimates (see e.g.,
\citealt{FT86,SFvP87,OGP09,GOC10,GO13}).
In addition the hard state bursts of 4U 1820--30, which occur above the
$\sim\!3$ per cent of Eddington threshold, have been used to make 
mass-radius constraints \citep{vPL87,GWC10}.
However, in both cases the data are not consistent with the NS atmosphere
model predictions (see also the critical discussion in \citealt{GZM13}).
There are various studies about the NS equation of state that are at least
partly based on these measurements (e.g., \citealt{SLB10,OBG10,LS14}), which in
light of the results presented here need to be revisited.
The reason for this statement is simple and holds even if our interpretation of
the $\Rapp$ evolution in soft state bursts is incorrect; it can be seen in the
two bursts of 4U 1608--52 shown in Figure \ref{fig:burst}(b) and (d).
Clearly, the hard state burst (panel b) is consistent with the atmosphere model
predictions, whereas the soft state burst (panel d) is not.
The results presented in Figure \ref{fig:K2K1} also argue that the soft state
bursts are not consistent with NS atmosphere models.
Previous NS mass-radius constraints are typically obtained from the soft state
bursts, using the NS atmosphere models by taking $\fc \approx 1.4$ in the
cooling tail, even if the predictions are not consistent with the data closer
to Eddington luminosity.
In addition, the hard state bursts are typically excluded from the analysis
using ad hoc arguments.
This is clearly inconsistent and these `selection effects' -- and their
impact on the NS mass-radius estimates -- are further highlighted for 4U
1608--52 in \citet{Poutanen14}.

One clear conclusion comes out of the results presented in this paper.
In order to make self-consistent NS mass-radius constraints using PRE-bursts,
one must choose only the hard state bursts that occur at persistent flux levels
below $\sim\!3$ per cent of Eddington.
However, further improvements into the NS atmosphere models are needed
before accurate mass-radius constraints can be obtained.
The stellar rotation is not yet incorporated to the \citet{SPW12} models, but
these models are being computed (Suleimanov et al., in prep.) and they will be
applied to the hard state bursts presented in this study (these results will be
presented in N{\"a}ttil{\"a} et al., in prep.).

\subsection{Open issues}

There are still many unanswered issues in the rich \xte/PCA data set regarding
the behaviour of the bursters.
The hard state bursts in the ultra-compact systems, 4U 1728--34 and 4U
1820--30, seem to behave differently with respect to the bursts from the rest
of the atoll sources in the sample; the origin for this is not understood at
present. It would be interesting to investigate if the behaviour of other
ultra-compact systems (see e.g., \citealt{intZJM07}) are also different from the
atolls with longer orbital periods. However, the $K$-ratios cannot be easily
computed for most of the candidate systems because they either do not show
PRE-bursts or the bursts are too faint to resolve the cooling track from
$F_{\rm td}$ to $F_{\rm td/2}$. A careful comparison with the Z-sources might
also be useful. For example, the soft state bursts of GX 17+2 follow the
canonical $L \propto T^4$ track, which implies $K$-ratios of $K_{\rm td/2} /
K_{\rm td} \approx 1$ (see e.g., \citealt{KHVdK02}, figure 13), but, as
discussed in Section 2, the intense persistent emission complicates the analysis
substantially. 

In Figure \ref{fig:K2K1} one can also see various hard state bursts with high
$K$-ratios of $K_{\rm td/2} / K_{\rm td} \gtrsim 3.6$. These bursts tend to also
be the longest and the most energetic ones. One of the systems showing these
bursts is 4U 1608--52 and, as in \citet{Poutanen14}, we
speculate that in these bursts the nuclear burning ashes might
have reached the photosphere \citep{WBS06}, leading to a lower $\fc$ at around
half Eddington luminosity, and thus higher $K$-ratio.

In addition, from Figure \ref{fig:K2K1} it is evident that there are a few
interesting `non-hard-state' bursts that have $K$-ratios consistent with the
\citet{SPW12} models.
Especially the B\#12 from 4U 1636--536 that is highlighted in Figure
\ref{fig:K2K1} seems to be a clear outlier. 
The touchdown flux of $\Ftd \! = \! (0.409 \pm 0.009) \times
10^{-7}\,\ergcm2s$ for this particular burst is clearly below the mean of 4U
1636--536 $\Ftdmean\! = \! 0.69 \times 10^{-7}\,\ergcm2s$. 
It was argued by \citet{GPM06} that for this faint PRE burst, the Eddington
limit for hydrogen-rich atmosphere was reached, whereas the rest of the PRE
bursts reach the Eddington limit for a pure helium atmosphere.
However, it is not clear why the presence of the hydrogen layer in this burst
causes the $K$-ratio to be so radically different from the rest of the soft state
bursts.
By comparing this faint PRE burst to the bursts from other sources in the sample
that show similar bi-modality in the $\Ftd$ values, we can see very interesting
differences between the sources:
\begin{itemize}
  \item [-] The PRE bursts in the faint $\Ftd$ group of 4U 1608--52 have high
$K$-ratios, likely because they all occur in the hard state below the $\sim\!3$
per cent of Eddington threshold.
  \item [-] In contrast to 4U 1608--52, the only hard state PRE burst of Aql
X--1 (B\#9) is clearly in the bright $\Ftd$ group, but it still has a high
$K$-ratio. On the other hand, the soft state bursts of Aql X--1 have $K$-ratios
of the order of unity with no apparent dependence between $\Ftd$ (nor $\Fper$),
indicating that the lower $\Ftd$ value in the B\#12 from 4U 1636--536 was
not the only factor leading to a high $K$-ratio.
  \item [-] Also, the faint PRE bursts of 4U 1724--307 (B\#2) and HETE
J1900.1--2455 (B\#5) both occur in the soft state and at highest $\Fper$,
but they have $K_{\rm td/2} / K_{\rm td}\! \approx \! 1.0$.
  \item [-] Finally, it is also interesting to note that there is absolutely
nothing remarkable in the only faint PRE burst of 4U 1735--44 (B\#4) in
terms of $\Fper$ or its position in the CC-diagram.
\end{itemize}
The cause for these differences cannot be related to the stellar spin,
because both 4U 1608--52 and Aql X--1 are fast rotators with $f_{\rm spin}\!
\sim\! 600$ Hz. Magnetic fields are likely not important either in this respect,
because both Aql X--1 and HETE J1900.1--2455 are intermittent pulsars, but still
show clear differences in their bursting behaviours. The persistent flux level
should not affect either, because the faint PRE burst of 4U 1735--44 (B\#4)
takes place at comparable $\Mdot$ as the B\#12 from 4U 1636--536, but, as
noted above, it has $K_{\rm td/2} / K_{\rm td}\! \approx\! 1.0$. Perhaps the
determining factor is the composition of the accreted fuel, as the ultra-compact
binary 4U 1820--30 stands out again as having the stable $\Ftd$ values.
Or, alternatively, some favourable combination of these factors might
occasionally cause the accretion flow to disturb the NS photosphere to a lesser
extent, thus leading to $K$-ratios which are consistent with the \citet{SPW12}
model predictions. But it must be admitted that it is very hard to draw solid
conclusions of this diversity based on the limited sample of bursters we have
studied. 

The inner hot flow is optically thin in the low-luminosity, hard states and if
the disc truncation radius is large enough, we probably always see the entire
NS surface during X-ray bursts in this state. 
However, the hot flow might also leave its imprint on the observed burst
spectra.
We have seen that some bursters show a large scatter in the absolute $\Rbb$
values between different hard state bursts (see also \citealt{GPO12}), although
they show the expected increase of $\Rbb$ values in the initial cooling phases.
This effect is not that well visible in the $K$-ratios that are shown in Figure
\ref{fig:K2K1}, because the individual X-ray bursts tend to follow the NS
atmosphere model predictions and thus by taking the ratio of the two $\Rbb$
values, the scatter is cancelled out to some extent.
This scatter is especially prominent for the two bursters showing large number
of bursts: 4U 1636--536 \citep{ZMA11} and 4U 1728--34 \citep{GPO12}, but similar
variations are also in seen other systems, like 4U 1608--52 \citep{Poutanen14}.
It is possible that a fraction of burst photons are up-scattered by the
energetic electrons in the hot flow producing a high-energy tail (and at the
same time cooling the electrons) thus altering the spectrum of the persistent
emission \citep{JZC14}.
This effect might cause extra spectral hardening of the burst emission and
influence $\fc$, and if the properties of the hot flow vary from burst to
burst (as seen in the CC-diagrams), then it might cause the scatter in the
absolute $\Rbb$ values.
One cannot either rule out that a small, but variable amount of nuclear
burning ashes reaches the photosphere in all PRE-bursts (as in the $K_{\rm td/2}
/ K_{\rm td} \gtrsim 3.6$ bursts), leading to burst-to-burst $\fc$ variations,
which
also might cause the observed $\Rbb$ scatter.
However, these possibilities should not be as prominent sources of systematic
error in the mass-radius estimates as the spreading layer in the high persistent
flux, soft state bursts.
Furthermore, the study by \citet{WGP13} indicates that the burst spectra
are better described statistically if the persistent emission is allowed to vary
by letting the `background' model normalisation vary by a factor $f_{\rm a}$,
where high $f_{\rm a}$ values are interpreted as an increase of the mass
accretion rate during the burst.
However, it is not yet clear if the simple method used by \citet{WGP13}
adequately captures the complexity of burst-disc interactions. 
Indeed, \citet{POB14} recently argued that the presence of high frequency
quasi-periodic oscillations (QPO) before and during the X-ray bursts, while
$f_{\rm a}$ is still above unity, is not consistent with this methodology: 
the high frequency QPOs are thought to be produced in the inner disc regions,
but the high $f_{\rm a}$ values imply that this region would be accreted onto
the NS.
However, this argument breaks down if the QPOs are associated with the
spreading layer as suggested by the Fourier-frequency resolved spectra
\citep{GRM03}.

These are clearly important issues that should be investigated further
because they might affect the NS mass-radius measurements.
This means that the applicability of the new atmosphere models -- and the
`cooling tail method' in general \citep{SPR11} -- for all the hard state
bursts is, therefore, not yet fully clear.
These issues are, however, under investigation and they will be addressed in
future publications.

\section{Summary and conclusions}

We have presented an analysis of 246 X-ray bursts from 11 LMXB systems.
Our main finding is that the evolution of the black body radii during the
initial cooling phases of X-ray bursts depends on the spectral state and the
accretion rate.
These differences are seen in the $K$-ratios, i.e., in the ratio between the
black body normalisation at `half-touchdown flux' and at the touchdown $K_{\rm
td/2} / K_{\rm td}$.
The hard state bursts tend show a characteristic increase of the black body
radius in the early cooling phase ($K_{\rm td/2} / K_{\rm td} > 2$), which is
consistent with NS atmosphere model predictions of \citet{SPW12}.
In particular, we find that only hard state bursts that have persistent fluxes
below a critical threshold of about $\sim\!3$ per cent of the Eddington value 
follow the NS atmosphere model predictions. 
Practically all soft state bursts and even hard state bursts from the candidate
ultra-compact systems that take place above this critical threshold do not; they
instead have $K_{\rm td/2} / K_{\rm td} \approx 1$, which is not consistent with
any NS atmosphere model prediction.

This result supports the interpretation of \citet{SPR11}, where bursts of
4U 1724--307 were analysed; the accretion flow plays an important role in
shaping the energy spectra of X-ray bursts. 
In the soft state the spreading layer that engulfs the NS causes the colour
correction factor to have a high, constant value of about $\fc \! \approx \! 1.6
- 1.8$ over a large range of luminosities from the Eddington flux down to a
fraction of it.
In addition, the spreading layer can also block part of the NS
surface from view in the soft state, but not necessarily in the hard state.
These two effects together can explain why the black body radii are constant
in the soft state bursts and why large differences are seen in the inferred NS
radii between hard- and soft state bursts.
The results presented here show that this behaviour is ubiquitous among atoll
sources.

The effects of the accretion flow are ignored in almost all previous studies
that use the soft state bursts to constrain NS masses and radii.
Our results indicate that the X-ray bursts in this state are influenced by the
accretion flow to such an extent that all previous NS mass-radius estimates need
to be revised. 
This is because the two main assumptions that are used -- i.e. $\fc=1.4$ and 
the entire NS surface is seen -- are not consistent with the data.
In fact, disentangling the effects of the accretion flow from the burst
emission might be so non-unique and non-trivial that the soft state bursts
should not be used to constrain NS masses and radii at all.
Rather, they might be better laboratories to investigate the dynamical behaviour
of spreading layers.

\section*{Acknowledgements}

We thank the referees for their comments and suggestions that helped to 
improve and clarify this paper.
JJEK acknowledges partial financial support from the Finnish Graduate School in
Astronomy and Space Physics, the Emil Aaltonen Foundation and the
V\"{a}is\"{a}l\"{a} Foundation.
JN also acknowledges financial support from the V\"{a}is\"{a}l\"{a} Foundation.
This research was supported by the Academy of Finland grant 268740 (JP). 
VS thanks DFG (grant SFB/Transregio 7 ``Gravitational Wave Astronomy''), Russian
Foundation of Fundamental Research (grant 12-02-97006-r-povolzhe-a) and the
COST Action MP1104 for support.
MR is supported by the grant of Russian Scientific Foundation RNF 14-12-01287.
DKG is the recipient of an Australian Research Council Future Fellowship
(project FT0991598).
We thank the International Space Science Institute (ISSI) located in Bern,
Switzerland, for sponsoring an International Team on type-I X-ray bursts where
early results of this project were discussed.
This research made use of the NASA Astrophysics Data System and of the data
obtained from the High Energy Astrophysics Science Archive (HEASARC), which is a
service of the Astrophysics Science Division at NASA/GSFC and the High Energy
Astrophysics Division of the Smithsonian Astrophysical Observatory. 

\bibliographystyle{mn2e}   
%\bibliography{mnemonic,lmxb}

\label{lastpage}

\appendix

\section{A Catalogue of PRE-bursts used in this study}

\onecolumn

\begin{table*}
\begin{minipage}{170mm}
\caption{
A catalogue of PRE-bursts used in this study. The B\# is the burst ID
number used in this work and the bracketed (G8\#) ID number used in 
\citet{GMH08} is also given for easier cross comparison. The touchdown fluxes
$\Ftd$ are in units of $10^{-7}\,{\rm erg\,cm^{-2}\,s^{-1}}$, whereas the
persistent flux is given in $10^{-9}\, {\rm erg\,cm^{-2}\,s^{-1}}$ ($1\sigma$
errors). HC and SC denote the hard- and the soft colours, respectively. The
symbol $^a$ highlights the three differences with respect to the \citet{GMH08}
catalogue: the B\#47 of 4U 1636--536 and the B\#42 of 4U 1728--34 were
not classified as PRE bursts by \citet{GMH08} and the B\#34 of 4U 1728--34
was not in their catalogue at all.}
\label{table:bigtable}
\begin{tabular}{lccccccc}
\hline\hline

B\# (G8\#) & OBSID & Date (MJD) & $\Ftd$ & $K_{\rm td/2} / K_{\rm td}$ & HC &
SC & $\Fper$\\
\hline
\multicolumn{8}{c}{4U 1608--52} \\
\hline
 1 (5) & 30062-01-01-00 &  50899.587729 & $ 1.87 \pm 0.02$ & $ 1.08 \pm
0.05$ & $ 0.605 \pm 0.008$ & $ 1.45 \pm 0.07$ & $ 3.03 \pm
0.07$ \\ 
 2 (8) & 30062-01-02-05 &  50914.275359 & $ 1.220 \pm 0.015$ & $ 3.2 \pm
0.2$ & $ 0.669 \pm 0.014$ & $ 1.54 \pm 0.03$ & $ 0.933 \pm
0.010$ \\ 
 3 (9) & 50052-02-01-01 &  51612.031543 & $ 1.39 \pm 0.03$ & $ 1.7 \pm
0.3$ & $ 0.57 \pm 0.05$ & $ 1.4 \pm 0.7$ & $ 1.7 \pm
0.5$ \\ 
 4 (10) & 50052-01-04-00 &  51614.071954 & $ 1.19 \pm 0.02$ & $ 3.8
\pm 0.2$ & $ 0.91 \pm 0.02$ & $ 1.66 \pm 0.04$ & $ 0.732 \pm
0.010$ \\ 
 5 (21) & 70059-01-20-00 &  52524.102263 & $ 1.77 \pm 0.03$ & $ 1.06
\pm 0.07$ & $ 0.65 \pm 0.03$ & $ 1.3 \pm 0.2$ & $ 6.3 \pm
0.7$ \\ 
 6 (22) & 70059-01-21-00 &  52526.160753 & $ 1.48 \pm 0.02$ & $ 1.01
\pm 0.05$ & $ 0.635 \pm 0.008$ & $ 1.55 \pm 0.09$ & $ 7.2 \pm
0.2$ \\ 
 7 (23) & 70059-03-01-00 &  52529.180035 & $ 1.67 \pm 0.06$ & $ 1.04
\pm 0.10$ & $ 0.651 \pm 0.008$ & $ 1.54 \pm 0.03$ & $ 4.30 \pm
0.04$ \\ 
 8 (25) & 70058-01-39-00 &  52536.318988 & $ 1.61 \pm 0.02$ & $ 1.58
\pm 0.07$ & $ 0.635 \pm 0.014$ & $ 1.48 \pm 0.15$ & $ 2.63 \pm
0.13$ \\ 
 9 (26) & 70069-01-01-00 &  52542.501479 & $ 1.27 \pm 0.02$ & $ 3.8
\pm 0.2$ & $ 0.98 \pm 0.02$ & $ 1.38 \pm 0.06$ & $ 0.754 \pm
0.012$ \\ 
10 (27) & 70059-01-26-00 &  52546.900131 & $ 1.21 \pm 0.03$ & $ 3.3
\pm 0.2$ & $ 0.74 \pm 0.02$ & $ 1.60 \pm 0.05$ & $ 0.729 \pm
0.011$ \\ 
11 (31) & 90408-01-04-04 &  53104.408629 & $ 1.21 \pm 0.05$ & $ 3.4
\pm 0.3$ & $ 1.14 \pm 0.02$ & $ 1.56 \pm 0.06$ & $ 0.935 \pm
0.011$ \\ 
12  & 93408-01-23-02 &  54434.974037 & $ 1.65 \pm 0.04$ & $ 1.34 \pm
0.11$ & $ 0.636 \pm 0.012$ & $ 1.57 \pm 0.07$ & $ 3.34 \pm
0.08$ \\ 
13  & 93408-01-25-06 &  54452.116158 & $ 1.03 \pm 0.03$ & $ 3.9 \pm
0.3$ & $ 1.252 \pm 0.015$ & $ 1.69 \pm 0.04$ & $ 1.748 \pm
0.013$ \\ 
14  & 93408-01-26-04 &  54461.031219 & $ 1.12 \pm 0.02$ & $ 4.9 \pm
0.3$ & $ 1.29 \pm 0.04$ & $ 1.8 \pm 0.3$ & $ 1.66 \pm
0.08$ \\ 
15  & 93408-01-59-03 &  54692.075270 & $ 1.18 \pm 0.03$ & $ 3.9 \pm
0.3$ & $ 0.90 \pm 0.02$ & $ 1.28 \pm 0.06$ & $ 0.83 \pm
0.02$ \\ 
16  & 95334-01-03-08 &  55270.220872 & $ 1.62 \pm 0.04$ & $ 1.32 \pm
0.11$ & $ 0.628 \pm 0.014$ & $ 1.4 \pm 0.2$ & $ 6.9 \pm
0.4$ \\ 
17  & 96423-01-11-01 &  55725.155733 & $ 1.59 \pm 0.03$ & $ 0.94 \pm
0.06$ & $ 0.705 \pm 0.010$ & $ 1.55 \pm 0.08$ & $ 4.98 \pm
0.12$ \\ 
18  & 96423-01-35-00 &  55890.371280 & $ 1.08 \pm 0.02$ & $ 4.2 \pm
0.4$ & $ 1.42 \pm 0.04$ & $ 1.86 \pm 0.15$ & $ 1.56 \pm
0.03$ \\ 
\hline
\multicolumn{8}{c}{4U 1636--536} \\
\hline
 1 (1) & 10088-01-07-02 &  50445.944732 & $ 0.714 \pm 0.011$ & $ 1.06 \pm
0.07$ & $ 0.708 \pm 0.006$ & $ 1.56 \pm 0.06$ & $ 4.42 \pm
0.08$ \\ 
 2 (3) & 10088-01-08-01 &  50446.977634 & $ 0.604 \pm 0.010$ & $ 0.87 \pm
0.06$ & $ 0.703 \pm 0.007$ & $ 1.55 \pm 0.06$ & $ 4.07 \pm
0.07$ \\ 
 3 (4) & 10088-01-08-03 &  50448.734636 & $ 0.681 \pm 0.011$ & $ 1.43 \pm
0.10$ & $ 0.726 \pm 0.008$ & $ 1.52 \pm 0.06$ & $ 4.41 \pm
0.08$ \\ 
 4 (6) & 30053-02-02-02 &  51044.490033 & $ 0.750 \pm 0.013$ & $ 0.92 \pm
0.07$ & $ 0.725 \pm 0.009$ & $ 1.42 \pm 0.08$ & $ 4.02 \pm
0.12$ \\ 
 5 (7) & 30053-02-01-02 &  51045.153577 & $ 0.77 \pm 0.02$ & $ 1.23 \pm
0.09$ & $ 0.728 \pm 0.009$ & $ 1.45 \pm 0.08$ & $ 4.03 \pm
0.10$ \\ 
 6 (9) & 40028-01-02-00 &  51236.367009 & $ 0.707 \pm 0.012$ & $ 1.11 \pm
0.08$ & $ 0.702 \pm 0.009$ & $ 1.47 \pm 0.06$ & $ 3.81 \pm
0.08$ \\ 
 7 (10) & 40028-01-04-00 &  51297.072674 & $ 0.689 \pm 0.014$ & $ 1.12
\pm 0.09$ & $ 0.674 \pm 0.006$ & $ 1.15 \pm 0.06$ & $ 5.6 \pm
0.3$ \\ 
 8 (12) & 40028-01-06-00 &  51339.247576 & $ 0.683 \pm 0.012$ & $ 0.85
\pm 0.06$ & $ 0.717 \pm 0.014$ & $ 1.37 \pm 0.13$ & $ 3.4 \pm
0.2$ \\ 
 9 (13) & 40028-01-08-00 &  51347.988936 & $ 0.563 \pm 0.015$ & $ 1.48
\pm 0.14$ & $ 0.718 \pm 0.008$ & $ 1.75 \pm 0.09$ & $ 5.04 \pm
0.11$ \\ 
10 (14) & 40030-03-04-00 &  51348.730536 & $ 0.675 \pm 0.012$ & $ 1.19
\pm 0.08$ & $ 0.707 \pm 0.014$ & $ 1.50 \pm 0.14$ & $ 3.7 \pm
0.2$ \\ 
11 (15) & 40031-01-01-06 &  51350.796442 & $ 0.654 \pm 0.012$ & $ 1.26
\pm 0.08$ & $ 0.720 \pm 0.013$ & $ 1.46 \pm 0.12$ & $ 3.69 \pm
0.15$ \\ 
12 (16) & 40028-01-10-00 &  51446.862370 & $ 0.409 \pm 0.009$ & $ 2.0
\pm 0.2$ & $ 0.771 \pm 0.011$ & $ 1.67 \pm 0.12$ & $ 7.8 \pm
0.2$ \\ 
13 (20) & 40028-01-15-00 &  51710.213022 & $ 0.79 \pm 0.02$ & $ 0.68
\pm 0.05$ & $ 0.729 \pm 0.009$ & $ 1.61 \pm 0.08$ & $ 4.05 \pm
0.09$ \\ 
14 (21) & 40028-01-18-00 &  51765.055325 & $ 0.637 \pm 0.014$ & $ 1.08
\pm 0.09$ & $ 0.738 \pm 0.009$ & $ 1.56 \pm 0.08$ & $ 3.84 \pm
0.09$ \\ 
15 (22) & 40028-01-18-00 &  51765.373534 & $ 0.75 \pm 0.03$ & $ 1.25
\pm 0.15$ & $ 0.727 \pm 0.009$ & $ 1.58 \pm 0.07$ & $ 3.63 \pm
0.08$ \\ 
16 (23) & 40028-01-19-00 &  51768.981502 & $ 0.765 \pm 0.012$ & $ 0.79
\pm 0.05$ & $ 0.731 \pm 0.012$ & $ 1.48 \pm 0.12$ & $ 3.59 \pm
0.14$ \\ 
17 (24) & 40028-01-20-00 &  51820.981815 & $ 0.723 \pm 0.012$ & $ 1.02
\pm 0.07$ & $ 0.723 \pm 0.007$ & $ 1.59 \pm 0.04$ & $ 3.37 \pm
0.04$ \\ 
18 (25) & 50030-02-01-00 &  51853.182628 & $ 0.579 \pm 0.011$ & $ 1.39
\pm 0.10$ & $ 0.698 \pm 0.009$ & $ 1.59 \pm 0.10$ & $ 4.59 \pm
0.13$ \\ 
19 (26) & 50030-02-02-00 &  51860.752408 & $ 0.609 \pm 0.011$ & $ 1.09
\pm 0.08$ & $ 0.705 \pm 0.007$ & $ 1.76 \pm 0.06$ & $ 4.89 \pm
0.07$ \\ 
20 (27) & 50030-02-04-00 &  51937.116815 & $ 0.63 \pm 0.02$ & $ 1.16
\pm 0.10$ & $ 0.723 \pm 0.010$ & $ 1.52 \pm 0.09$ & $ 3.79 \pm
0.11$ \\ 
21 (28) & 50030-02-05-01 &  51941.876280 & $ 0.637 \pm 0.011$ & $ 1.08
\pm 0.08$ & $ 0.69 \pm 0.02$ & $ 1.5 \pm 0.2$ & $ 4.3 \pm
0.2$ \\ 
22 (29) & 50030-02-05-00 &  51942.100935 & $ 0.68 \pm 0.02$ & $ 1.32
\pm 0.11$ & $ 0.720 \pm 0.012$ & $ 1.45 \pm 0.12$ & $ 4.1 \pm
0.2$ \\ 
23 (30) & 50030-02-09-00 &  52004.713947 & $ 1.06 \pm 0.02$ & $ 1.18
\pm 0.09$ & $ 0.752 \pm 0.013$ & $ 1.54 \pm 0.08$ & $ 3.68 \pm
0.09$ \\ 
24 (31) & 50030-02-10-00 &  52029.228869 & $ 0.76 \pm 0.03$ & $ 0.99
\pm 0.10$ & $ 0.712 \pm 0.015$ & $ 1.46 \pm 0.14$ & $ 2.76 \pm
0.13$ \\ 
25 (34) & 60032-01-02-00 &  52075.135469 & $ 0.76 \pm 0.02$ & $ 1.35
\pm 0.12$ & $ 0.758 \pm 0.012$ & $ 1.60 \pm 0.05$ & $ 2.07 \pm
0.03$ \\ 
26 (38) & 60032-01-06-01 &  52149.279399 & $ 0.647 \pm 0.011$ & $ 1.13
\pm 0.08$ & $ 0.724 \pm 0.010$ & $ 1.60 \pm 0.08$ & $ 3.13 \pm
0.07$ \\
27 (45) & 60032-01-12-00 &  52182.616870 & $ 0.781 \pm 0.015$ & $ 1.15
\pm 0.08$ & $ 0.71 \pm 0.02$ & $ 1.52 \pm 0.14$ & $ 2.60 \pm
0.11$ \\ 
28 (49) & 60032-01-14-01 &  52214.318968 & $ 0.701 \pm 0.014$ & $ 1.12
\pm 0.09$ & $ 0.73 \pm 0.02$ & $ 1.5 \pm 0.2$ & $ 3.2 \pm
0.2$ \\ 
29 (50) & 60032-01-18-00 &  52273.691505 & $ 0.541 \pm 0.011$ & $ 2.07
\pm 0.14$ & $ 0.82 \pm 0.02$ & $ 1.60 \pm 0.11$ & $ 1.79 \pm
0.06$ \\ 
30 (61) & 60032-01-20-00 &  52283.019196 & $ 0.84 \pm 0.02$ & $ 1.13
\pm 0.08$ & $ 0.778 \pm 0.010$ & $ 1.60 \pm 0.02$ & $ 2.08 \pm
0.02$ \\ 
31 (62) & 60032-01-20-01 &  52283.534309 & $ 0.736 \pm 0.014$ & $ 0.98
\pm 0.07$ & $ 0.731 \pm 0.015$ & $ 1.60 \pm 0.10$ & $ 2.15 \pm
0.06$ \\ 
32 (68) & 60032-05-01-00 &  52286.054732 & $ 0.599 \pm 0.010$ & $ 1.68
\pm 0.11$ & $ 0.937 \pm 0.012$ & $ 1.67 \pm 0.03$ & $ 1.52 \pm
0.02$ \\
33 (72) & 60032-05-02-00 &  52286.555354 & $ 0.686 \pm 0.014$ & $ 1.28
\pm 0.10$ & $ 0.936 \pm 0.014$ & $ 1.63 \pm 0.03$ & $ 1.65 \pm
0.02$ \\ 
34 (79) & 60032-05-04-00 &  52287.522590 & $ 0.541 \pm 0.009$ & $ 1.72
\pm 0.12$ & $ 0.939 \pm 0.013$ & $ 1.65 \pm 0.02$ & $ 1.611 \pm
0.015$ \\ 
\hline\hline\\
\end{tabular}
\end{minipage}
\end{table*}

\begin{table*}
\begin{minipage}{170mm}
\contcaption{}
\label{bigtable2}
\begin{tabular}{lccccccc}
\hline\hline

B\# (G8\#) & OBSID & Date (MJD) & $\Ftd$ & $K_{\rm td/2} / K_{\rm td}$ & HC &
SC & $\Fper$\\
\hline
35 (86) & 60032-05-06-00 &  52288.515004 & $ 0.80 \pm 0.02$ & $ 1.23
\pm
0.11$ & $ 0.804 \pm 0.011$ & $ 1.61 \pm 0.02$ & $ 1.896 \pm
0.015$ \\ 
36 (87) & 60032-05-07-00 &  52288.975073 & $ 0.632 \pm 0.008$ & $ 1.84
\pm
0.10$ & $ 0.873 \pm 0.011$ & $ 1.57 \pm 0.04$ & $ 1.69 \pm
0.02$ \\ 
37 (88) & 60032-05-07-01 &  52289.293513 & $ 0.568 \pm 0.011$ & $ 1.94
\pm
0.14$ & $ 0.896 \pm 0.013$ & $ 1.61 \pm 0.04$ & $ 1.60 \pm
0.02$ \\ 
38 (94) & 60032-05-09-00 &  52289.977634 & $ 0.663 \pm 0.015$ & $ 1.53
\pm
0.11$ & $ 0.841 \pm 0.010$ & $ 1.56 \pm 0.04$ & $ 1.75 \pm
0.02$ \\ 
39 (109) & 60032-05-12-00 &  52304.963829 & $ 0.629 \pm 0.011$ & $ 1.02
\pm
0.07$ & $ 0.669 \pm 0.014$ & $ 1.14 \pm 0.14$ & $ 5.3 \pm
0.4$ \\ 
40 (110) & 60032-05-13-00 &  52310.932538 & $ 0.72 \pm 0.03$ & $ 0.88
\pm
0.10$ & $ 0.719 \pm 0.010$ & $ 1.67 \pm 0.10$ & $ 4.86 \pm
0.12$ \\ 
41 (111) & 60032-05-14-00 &  52316.733415 & $ 0.71 \pm 0.02$ & $ 1.39
\pm 0.14$ & $ 0.705 \pm 0.007$ & $ 1.68 \pm 0.05$ & $ 4.45 \pm
0.06$ \\ 
42 (115) & 60032-05-18-00 &  52390.214092 & $ 0.561 \pm 0.012$ & $ 1.39
\pm 0.13$ & $ 0.75 \pm 0.02$ & $ 1.7 \pm 0.2$ & $ 3.0 \pm
0.2$ \\ 
43 (122) & 60032-05-22-00 &  52551.251902 & $ 0.629 \pm 0.013$ & $ 1.53
\pm 0.12$ & $ 0.811 \pm 0.012$ & $ 1.59 \pm 0.08$ & $ 3.08 \pm
0.08$ \\ 
44 (125) & 80425-01-01-00 &  52899.945004 & $ 0.665 \pm 0.014$ & $ 1.82
\pm 0.13$ & $ 0.96 \pm 0.02$ & $ 1.60 \pm 0.05$ & $ 1.48 \pm
0.02$ \\ 
45 (136) & 91024-01-42-00 &  53516.313817 & $ 0.75 \pm 0.02$ & $ 1.14
\pm 0.10$ & $ 0.732 \pm 0.012$ & $ 1.59 \pm 0.06$ & $ 2.23 \pm
0.04$ \\ 
46 (137) & 91024-01-46-00 &  53524.389520 & $ 0.694 \pm 0.011$ & $ 1.22
\pm 0.08$ & $ 0.742 \pm 0.009$ & $ 1.62 \pm 0.05$ & $ 2.51 \pm
0.04$ \\ 
47 (148)$^a$ & 91024-01-80-00 &  53592.234454 & $ 0.726 \pm 0.012$ & $
1.18 \pm 0.08$ & $ 0.725 \pm 0.006$ & $ 1.58 \pm 0.02$ & $
3.67 \pm 0.03$ \\ 
48 (149) & 91024-01-82-00 &  53596.088511 & $ 0.65 \pm 0.02$ & $ 1.28
\pm 0.13$ & $ 0.693 \pm 0.012$ & $ 1.58 \pm 0.13$ & $ 4.9 \pm
0.2$ \\ 
49 (150) & 91024-01-83-00 &  53598.074040 & $ 0.71 \pm 0.02$ & $ 0.77
\pm 0.07$ & $ 0.710 \pm 0.010$ & $ 1.68 \pm 0.11$ & $ 4.83 \pm
0.13$ \\ 
50 (168) & 91024-01-30-10 &  53688.952608 & $ 0.79 \pm 0.02$ & $ 1.08
\pm 0.10$ & $ 0.711 \pm 0.011$ & $ 1.50 \pm 0.10$ & $ 4.08 \pm
0.13$ \\ 
51  & 91152-05-02-00 &  53919.074680 & $ 0.71 \pm 0.02$ & $ 1.21 \pm
0.11$ & $ 0.759 \pm 0.011$ & $ 1.47 \pm 0.11$ & $ 4.15 \pm
0.14$ \\ 
52  & 92023-01-72-00 &  53940.493218 & $ 0.694 \pm 0.014$ & $ 0.90 \pm
0.06$ & $ 0.718 \pm 0.012$ & $ 1.50 \pm 0.12$ & $ 3.26 \pm
0.12$ \\ 
53  & 92023-01-10-10 &  54012.556123 & $ 0.63 \pm 0.02$ & $ 1.48 \pm
0.14$ & $ 0.697 \pm 0.008$ & $ 1.71 \pm 0.03$ & $ 3.70 \pm
0.04$ \\ 
54  & 92023-01-29-10 &  54050.902738 & $ 0.72 \pm 0.02$ & $ 1.07 \pm
0.10$ & $ 0.72 \pm 0.02$ & $ 1.47 \pm 0.14$ & $ 2.37 \pm
0.11$ \\ 
55  & 92023-01-31-10 &  54054.249711 & $ 0.672 \pm 0.013$ & $ 1.12 \pm
0.09$ & $ 0.72 \pm 0.02$ & $ 1.5 \pm 0.2$ & $ 2.8 \pm
0.2$ \\ 
56  & 92023-01-60-10 &  54112.003181 & $ 0.76 \pm 0.02$ & $ 1.15 \pm
0.10$ & $ 0.73 \pm 0.02$ & $ 1.5 \pm 0.2$ & $ 2.56 \pm
0.13$ \\ 
57  & 92023-01-23-20 &  54222.420545 & $ 0.712 \pm 0.014$ & $ 1.18 \pm
0.09$ & $ 0.719 \pm 0.008$ & $ 1.56 \pm 0.06$ & $ 4.31 \pm
0.08$ \\ 
58  & 70036-01-02-01 &  54271.044500 & $ 0.786 \pm 0.015$ & $ 0.91 \pm
0.07$ & $ 0.719 \pm 0.014$ & $ 1.59 \pm 0.11$ & $ 2.77 \pm
0.09$ \\ 
59  & 70036-01-02-00 &  54272.092498 & $ 0.77 \pm 0.02$ & $ 0.96 \pm
0.08$ & $ 0.715 \pm 0.010$ & $ 1.57 \pm 0.05$ & $ 2.90 \pm
0.05$ \\ 
60  & 93091-01-01-00 &  54371.719668 & $ 0.711 \pm 0.014$ & $ 1.17 \pm
0.09$ & $ 0.777 \pm 0.013$ & $ 1.55 \pm 0.02$ & $ 1.81 \pm
0.02$ \\ 
61  & 93087-01-69-00 &  54416.318415 & $ 0.73 \pm 0.02$ & $ 1.30 \pm
0.12$ & $ 0.72 \pm 0.02$ & $ 1.5 \pm 0.2$ & $ 2.2 \pm
0.2$ \\ 
62  & 93087-01-24-10 &  54522.687075 & $ 0.78 \pm 0.03$ & $ 1.5 \pm
0.2$ & $ 0.79 \pm 0.03$ & $ 1.3 \pm 0.3$ & $ 2.3 \pm
0.2$ \\ 
63  & 93091-01-02-00 &  54523.579098 & $ 0.70 \pm 0.02$ & $ 1.4 \pm
0.2$ & $ 0.722 \pm 0.015$ & $ 1.63 \pm 0.08$ & $ 2.70 \pm
0.06$ \\ 
64  & 93087-01-28-10 &  54530.797894 & $ 0.65 \pm 0.02$ & $ 1.18 \pm
0.12$ & $ 0.731 \pm 0.014$ & $ 1.57 \pm 0.10$ & $ 2.49 \pm
0.08$ \\ 
65  & 93087-01-57-10 &  54588.165192 & $ 0.792 \pm 0.015$ & $ 0.96 \pm
0.07$ & $ 0.814 \pm 0.011$ & $ 1.53 \pm 0.02$ & $ 2.44 \pm
0.02$ \\ 
66  & 93087-01-70-10 &  54614.815982 & $ 0.68 \pm 0.02$ & $ 0.98 \pm
0.10$ & $ 0.71 \pm 0.02$ & $ 1.4 \pm 0.3$ & $ 3.1 \pm
0.3$ \\ 
67  & 93087-01-91-10 &  54656.604263 & $ 0.74 \pm 0.02$ & $ 1.02 \pm
0.10$ & $ 0.72 \pm 0.02$ & $ 1.6 \pm 0.3$ & $ 2.5 \pm
0.2$ \\ 
68  & 93087-01-04-20 &  54678.268528 & $ 0.76 \pm 0.02$ & $ 1.07 \pm
0.10$ & $ 0.770 \pm 0.014$ & $ 1.60 \pm 0.09$ & $ 2.32 \pm
0.06$ \\ 
69  & 94310-01-01-00 &  54904.833592 & $ 0.710 \pm 0.014$ & $ 1.16 \pm
0.09$ & $ 0.744 \pm 0.009$ & $ 1.59 \pm 0.03$ & $ 2.30 \pm
0.02$ \\ 
70  & 94310-01-03-00 &  55079.220357 & $ 0.67 \pm 0.02$ & $ 1.01 \pm
0.10$ & $ 0.72 \pm 0.02$ & $ 1.59 \pm 0.12$ & $ 2.16 \pm
0.08$ \\ 
71  & 94087-01-45-10 &  55110.235681 & $ 0.72 \pm 0.02$ & $ 1.2 \pm
0.2$ & $ 0.71 \pm 0.02$ & $ 1.66 \pm 0.11$ & $ 2.70 \pm
0.09$ \\ 
72  & 94087-01-73-10 &  55166.027634 & $ 0.598 \pm 0.015$ & $ 1.9 \pm
0.2$ & $ 0.702 \pm 0.014$ & $ 1.51 \pm 0.13$ & $ 3.73 \pm
0.15$ \\ 
73  & 94087-01-74-10 &  55168.322805 & $ 0.72 \pm 0.02$ & $ 0.94 \pm
0.09$ & $ 0.713 \pm 0.013$ & $ 1.61 \pm 0.12$ & $ 4.22 \pm
0.14$ \\ 
74  & 95087-01-39-00 &  55274.497261 & $ 0.64 \pm 0.02$ & $ 2.2 \pm
0.2$ & $ 0.790 \pm 0.015$ & $ 1.58 \pm 0.04$ & $ 1.59 \pm
0.02$ \\ 
75  & 95087-01-42-00 &  55280.541745 & $ 0.526 \pm 0.014$ & $ 1.4 \pm
0.2$ & $ 0.71 \pm 0.03$ & $ 1.6 \pm 0.3$ & $ 2.6 \pm
0.2$ \\ 
76  & 93082-06-06-00 &  55356.990762 & $ 0.70 \pm 0.02$ & $ 1.44 \pm
0.15$ & $ 0.71 \pm 0.02$ & $ 1.7 \pm 0.2$ & $ 5.0 \pm
0.2$ \\ 
77  & 95087-01-89-00 &  55374.685875 & $ 0.76 \pm 0.02$ & $ 2.0 \pm
0.2$ & $ 0.74 \pm 0.02$ & $ 1.5 \pm 0.2$ & $ 1.83 \pm
0.15$ \\ 
78  & 95087-01-01-10 &  55394.904737 & $ 0.65 \pm 0.02$ & $ 3.0 \pm
0.3$ & $ 0.80 \pm 0.02$ & $ 1.53 \pm 0.13$ & $ 1.51 \pm
0.06$ \\ 
79  & 95087-01-22-10 &  55436.152747 & $ 0.61 \pm 0.02$ & $ 1.39 \pm
0.15$ & $ 0.709 \pm 0.012$ & $ 1.71 \pm 0.10$ & $ 3.65 \pm
0.10$ \\ 
80  & 96087-01-46-00 &  55652.641696 & $ 0.74 \pm 0.03$ & $ 1.07 \pm
0.14$ & $ 0.711 \pm 0.015$ & $ 1.60 \pm 0.07$ & $ 2.76 \pm
0.06$ \\ 
81  & 96087-01-50-10 &  55857.000487 & $ 0.77 \pm 0.02$ & $ 1.18 \pm
0.10$ & $ 0.81 \pm 0.02$ & $ 1.57 \pm 0.13$ & $ 2.78 \pm
0.10$ \\ 
\hline
\multicolumn{8}{c}{4U 1702--429} \\
\hline
 1 (11) & 50025-01-01-00 &  51781.333039 & $ 0.813 \pm 0.015$ & $ 2.15
\pm 0.13$ & $ 1.003 \pm 0.012$ & $ 1.99 \pm 0.05$ & $ 1.37 \pm
0.02$ \\ 
 2 (19) & 80033-01-01-08 &  52957.629763 & $ 0.89 \pm 0.02$ & $ 2.4
\pm 0.2$ & $ 1.148 \pm 0.014$ & $ 1.57 \pm 0.04$ & $ 1.749 \pm
0.015$ \\ 
 3 (43) & 80033-01-19-04 &  53211.964665 & $ 0.80 \pm 0.02$ & $ 2.2
\pm 0.2$ & $ 1.01 \pm 0.03$ & $ 1.8 \pm 0.2$ & $ 1.52 \pm
0.05$ \\ 
 4 (44) & 80033-01-20-02 &  53212.794286 & $ 0.86 \pm 0.02$ & $ 2.21
\pm 0.15$ & $ 1.05 \pm 0.02$ & $ 1.97 \pm 0.09$ & $ 1.38 \pm
0.02$ \\ 
 5 (45) & 80033-01-21-00 &  53311.806086 & $ 0.76 \pm 0.02$ & $ 2.3
\pm 0.2$ & $ 0.92 \pm 0.04$ & $ 1.50 \pm 0.13$ & $ 1.38 \pm
0.04$ \\ 
\hline
\multicolumn{8}{c}{4U 1705--44} \\
\hline
 1 (1) & 20074-02-01-00 &  50495.947813 & $ 0.397 \pm 0.008$ & $ 1.17 \pm
0.09$ & $ 0.654 \pm 0.013$ & $ 1.7 \pm 0.2$ & $ 2.50 \pm
0.11$ \\ 
 2 (5) & 20073-04-01-03 &  50542.503568 & $ 0.401 \pm 0.008$ & $ 1.25 \pm
0.09$ & $ 0.670 \pm 0.008$ & $ 1.72 \pm 0.07$ & $ 2.58 \pm
0.04$ \\ 
 3 (21) & 40034-01-05-00 &  51333.396013 & $ 0.430 \pm 0.010$ & $ 1.04
\pm 0.10$ & $ 0.637 \pm 0.008$ & $ 1.86 \pm 0.06$ & $ 2.54 \pm
0.03$ \\ 
 4  & 93060-01-25-10 &  54764.450160 & $ 0.363 \pm 0.012$ & $ 1.6 \pm
0.2$ & $ 0.66 \pm 0.02$ & $ 1.7 \pm 0.2$ & $ 2.70 \pm
0.14$ \\ 
 5  & 93060-01-28-10 &  54776.944564 & $ 0.369 \pm 0.012$ & $ 1.8 \pm
0.2$ & $ 0.76 \pm 0.02$ & $ 1.8 \pm 0.2$ & $ 1.30 \pm
0.05$ \\
\hline\hline\\
\end{tabular}
\end{minipage}
\end{table*}

\begin{table*}
\begin{minipage}{170mm}
\contcaption{}
\label{bigtable3}
\begin{tabular}{lccccccc}
\hline\hline

B\# (G8\#) & OBSID & Date (MJD) & $\Ftd$ & $K_{\rm td/2} / K_{\rm td}$ & HC &
SC & $\Fper$\\
\hline
\multicolumn{8}{c}{4U 1724--307} \\
\hline
 1 (1) & 10090-01-01-02 &  50395.292725 & $ 0.624 \pm 0.011$ & $ 2.66 \pm
0.15$ & $ 1.200 \pm 0.013$ & $ 1.63 \pm 0.03$ & $ 1.408 \pm
0.009$ \\ 
 2 (2) & 80138-06-06-00 &  53058.402090 & $ 0.429 \pm 0.010$ & $ 1.06 \pm
0.10$ & $ 0.74 \pm 0.02$ & $ 1.9 \pm 0.2$ & $ 2.21 \pm
0.11$ \\ 
 3 (3) & 90058-06-02-00 &  53147.218979 & $ 0.603 \pm 0.012$ & $ 1.69 \pm
0.12$ & $ 0.83 \pm 0.02$ & $ 1.8 \pm 0.2$ & $ 1.36 \pm
0.05$ \\ 
 4  & 93044-06-04-00 &  54526.679905 & $ 0.65 \pm 0.02$ & $ 2.3 \pm
0.2$ & $ 1.22 \pm 0.02$ & $ 1.66 \pm 0.06$ & $ 1.511 \pm
0.015$ \\ 
\hline
\multicolumn{8}{c}{4U 1728--34} \\
\hline
 1 (8) & 10073-01-04-00 &  50131.895744 & $ 0.90 \pm 0.02$ & $ 1.36 \pm
0.10$ & $ 1.005 \pm 0.011$ & $ 1.88 \pm 0.08$ & $ 3.57 \pm
0.05$ \\ 
 2 (9) & 10073-01-06-00 &  50135.965429 & $ 1.06 \pm 0.02$ & $ 2.13 \pm
0.14$ & $ 1.107 \pm 0.009$ & $ 2.03 \pm 0.03$ & $ 2.77 \pm
0.02$ \\ 
 3 (10) & 10073-01-07-00 &  50137.241242 & $ 1.03 \pm 0.03$ & $ 2.1
\pm 0.2$ & $ 1.131 \pm 0.010$ & $ 2.05 \pm 0.04$ & $ 2.65 \pm
0.02$ \\ 
 4 (21) & 20083-01-04-01 &  50718.472324 & $ 0.91 \pm 0.02$ & $ 1.23
\pm 0.08$ & $ 0.933 \pm 0.012$ & $ 1.96 \pm 0.10$ & $ 3.12 \pm
0.05$ \\ 
 5 (22) & 20083-01-04-01 &  50718.663265 & $ 0.810 \pm 0.013$ & $ 1.13
\pm 0.06$ & $ 0.890 \pm 0.007$ & $ 1.99 \pm 0.04$ & $ 3.24 \pm
0.02$ \\ 
 6 (27) & 30042-03-01-00 &  51086.423580 & $ 1.01\pm 0.02$ & $ 1.34
\pm 0.15$ & $ 1.208 \pm 0.008$ & $ 2.09 \pm 0.03$ & $ 4.93 \pm
0.03$ \\ 
 7 (28) & 30042-03-03-01 &  51110.106607 & $ 0.96 \pm 0.02$ & $ 1.5
\pm 0.2$ & $ 1.216 \pm 0.008$ & $ 2.03 \pm 0.04$ & $ 4.44 \pm
0.03$ \\ 
 8 (29) & 30042-03-06-00 &  51118.158997 & $ 1.04 \pm 0.02$ & $ 1.4
\pm 0.2$ & $ 1.232 \pm 0.015$ & $ 2.4 \pm 0.2$ & $ 4.49 \pm
0.09$ \\ 
 9 (30) & 30042-03-06-00 &  51118.291829 & $ 0.98 \pm 0.02$ & $ 1.6
\pm 0.2$ & $ 1.232 \pm 0.009$ & $ 2.01 \pm 0.04$ & $ 4.85 \pm
0.04$ \\ 
10 (31) & 30042-03-07-01 &  51119.947246 & $ 0.99 \pm 0.02$ & $ 1.7
\pm 0.2$ & $ 1.218 \pm 0.010$ & $ 1.99 \pm 0.05$ & $ 4.77 \pm
0.04$ \\ 
11 (32) & 30042-03-07-00 &  51120.085727 & $ 1.00 \pm 0.02$ & $ 1.4
\pm 0.2$ & $ 1.19 \pm 0.02$ & $ 2.06 \pm 0.14$ & $ 4.88 \pm
0.08$ \\ 
12 (33) & 30042-03-10-00 &  51127.815762 & $ 0.94 \pm 0.02$ & $ 1.7
\pm 0.2$ & $ 1.183 \pm 0.008$ & $ 2.06 \pm 0.03$ & $ 4.98 \pm
0.03$ \\ 
13 (34) & 30042-03-11-00 &  51128.026800 & $ 0.98 \pm 0.02$ & $ 1.6
\pm 0.2$ & $ 1.173 \pm 0.007$ & $ 2.07 \pm 0.03$ & $ 4.90 \pm
0.03$ \\ 
14 (35) & 30042-03-12-00 &  51128.681711 & $ 0.95 \pm 0.02$ & $ 1.19
\pm 0.13$ & $ 1.158 \pm 0.008$ & $ 2.09 \pm 0.03$ & $ 4.89 \pm
0.03$ \\ 
15 (36) & 30042-03-12-00 &  51128.814665 & $ 0.95 \pm 0.02$ & $ 1.33
\pm 0.15$ & $ 1.149 \pm 0.008$ & $ 2.10 \pm 0.03$ & $ 4.85 \pm
0.03$ \\ 
16 (37) & 30042-03-13-00 &  51129.014549 & $ 0.94 \pm 0.02$ & $ 1.35
\pm 0.15$ & $ 1.11 \pm 0.04$ & $ 1.8 \pm 0.3$ & $ 5.3 \pm
0.3$ \\ 
17 (38) & 30042-03-14-00 &  51133.424633 & $ 0.92 \pm 0.02$ & $ 0.78
\pm 0.09$ & $ 0.974 \pm 0.007$ & $ 2.06 \pm 0.04$ & $ 5.01 \pm
0.03$ \\ 
18 (39) & 30042-03-15-00 &  51133.673684 & $ 0.88 \pm 0.02$ & $ 1.16
\pm 0.15$ & $ 0.995 \pm 0.006$ & $ 2.09 \pm 0.03$ & $ 4.81 \pm
0.03$ \\ 
19 (41) & 30042-03-17-00 &  51134.573024 & $ 0.96 \pm 0.02$ & $ 1.13
\pm 0.13$ & $ 1.113 \pm 0.007$ & $ 2.10 \pm 0.03$ & $ 4.26 \pm
0.03$ \\ 
20 (43) & 30042-03-20-00 &  51196.991395 & $ 1.06 \pm 0.02$ & $ 1.9
\pm 0.2$ & $ 1.216 \pm 0.009$ & $ 2.00 \pm 0.04$ & $ 3.69 \pm
0.03$ \\ 
21 (44) & 40033-06-01-00 &  51198.144072 & $ 1.09 \pm 0.03$ & $ 1.68
\pm 0.11$ & $ 1.200 \pm 0.012$ & $ 1.99 \pm 0.08$ & $ 3.95 \pm
0.03$ \\ 
22 (45) & 40033-06-02-00 &  51200.267845 & $ 0.96 \pm 0.02$ & $ 1.39
\pm 0.08$ & $ 1.171 \pm 0.008$ & $ 2.05 \pm 0.04$ & $ 4.21 \pm
0.03$ \\ 
23 (46) & 40033-06-02-01 &  51201.993076 & $ 0.89 \pm 0.02$ & $ 1.59
\pm 0.12$ & $ 1.122 \pm 0.008$ & $ 2.08 \pm 0.04$ & $ 4.32 \pm
0.03$ \\
24 (47) & 40033-06-02-02 &  51202.358973 & $ 0.86 \pm 0.02$ & $ 1.89
\pm 0.12$ & $ 1.126 \pm 0.008$ & $ 2.12 \pm 0.03$ & $ 4.29 \pm
0.03$ \\ 
25 (48) & 40033-06-02-03 &  51204.001838 & $ 0.89 \pm 0.02$ & $ 1.20
\pm 0.08$ & $ 1.043 \pm 0.007$ & $ 2.09 \pm 0.04$ & $ 4.43 \pm
0.03$ \\ 
26 (49) & 40033-06-02-03 &  51204.130600 & $ 0.87 \pm 0.02$ & $ 1.14
\pm 0.09$ & $ 1.038 \pm 0.007$ & $ 2.10 \pm 0.04$ & $ 4.45 \pm
0.03$ \\ 
27 (51) & 40033-06-02-05 &  51206.141372 & $ 0.880 \pm 0.014$ & $ 1.21
\pm 0.07$ & $ 1.051 \pm 0.010$ & $ 2.08 \pm 0.07$ & $ 4.13 \pm
0.03$ \\ 
28 (52) & 40033-06-03-01 &  51208.985938 & $ 1.02 \pm 0.02$ & $ 1.44
\pm 0.10$ & $ 1.053 \pm 0.008$ & $ 2.10 \pm 0.04$ & $ 3.76 \pm
0.03$ \\ 
29 (53) & 40033-06-03-02 &  51209.918759 & $ 0.80 \pm 0.02$ & $ 1.34
\pm 0.09$ & $ 0.935 \pm 0.013$ & $ 1.89 \pm 0.12$ & $ 4.14 \pm
0.09$ \\ 
30 (54) & 40033-06-03-02 &  51210.083146 & $ 0.81 \pm 0.02$ & $ 1.08
\pm 0.08$ & $ 0.948 \pm 0.006$ & $ 2.06 \pm 0.03$ & $ 4.06 \pm
0.02$ \\ 
31 (55) & 40033-06-03-05 &  51213.939185 & $ 0.864 \pm 0.013$ & $ 0.98
\pm 0.05$ & $ 0.929 \pm 0.014$ & $ 1.92 \pm 0.13$ & $ 3.76 \pm
0.08$ \\ 
32 (56) & 40027-06-01-00 &  51236.792781 & $ 0.93 \pm 0.02$ & $ 1.81
\pm 0.10$ & $ 1.091 \pm 0.009$ & $ 2.06 \pm 0.04$ & $ 2.67 \pm
0.02$ \\ 
33 (57) & 40027-06-01-02 &  51237.203189 & $ 0.87 \pm 0.02$ & $ 1.39
\pm 0.08$ & $ 1.086 \pm 0.009$ & $ 2.04 \pm 0.04$ & $ 2.69 \pm
0.02$ \\ 
34 (...) & 40027-06-01-06 &  51238.566977 & $ 0.92 \pm 0.02$ & $ 1.70
\pm 0.10$ & $ 1.094 \pm 0.011$ & $ 2.02 \pm 0.04$ & $ 2.75 \pm
0.02$ \\ 
35 (58) & 40027-06-01-03 &  51238.792243 & $ 0.99 \pm 0.02$ & $ 1.83
\pm 0.12$ & $ 1.125 \pm 0.011$ & $ 2.02 \pm 0.06$ & $ 2.75 \pm
0.02$ \\ 
36 (59) & 40027-06-01-08 &  51240.047298 & $ 0.919 \pm 0.015$ & $ 1.65
\pm 0.08$ & $ 1.100 \pm 0.009$ & $ 2.07 \pm 0.04$ & $ 2.89 \pm
0.02$ \\ 
37 (62) & 40027-08-01-01 &  51359.827252 & $ 0.93 \pm 0.02$ & $ 1.46
\pm 0.09$ & $ 1.087 \pm 0.012$ & $ 2.05 \pm 0.06$ & $ 2.89 \pm
0.03$ \\ 
38 (63) & 40027-08-03-00 &  51369.422561 & $ 1.07 \pm 0.02$ & $ 1.62
\pm 0.12$ & $ 1.156 \pm 0.012$ & $ 1.94 \pm 0.08$ & $ 2.99 \pm
0.04$ \\ 
39 (69) & 40019-03-01-00 &  51443.014309 & $ 0.83 \pm 0.02$ & $ 1.17
\pm 0.08$ & $ 0.957 \pm 0.010$ & $ 2.08 \pm 0.05$ & $ 3.14 \pm
0.03$ \\ 
40 (76) & 50029-23-02-01 &  51657.203959 & $ 0.93 \pm 0.02$ & $ 2.02
\pm 0.13$ & $ 1.107 \pm 0.009$ & $ 2.00 \pm 0.04$ & $ 2.54 \pm
0.02$ \\ 
41 (77) & 50029-23-02-02 &  51657.679170 & $ 1.04 \pm 0.02$ & $ 1.58
\pm 0.10$ & $ 1.094 \pm 0.010$ & $ 2.04 \pm 0.05$ & $ 2.56 \pm
0.02$ \\ 
42 (78)$^a$ & 50023-01-21-00 &  51691.713212 & $ 1.00 \pm 0.03$ & $
1.85 \pm 0.13$ & $ 1.179 \pm 0.015$ & $ 1.61 \pm 0.05$ & $
3.61 \pm 0.04$ \\ 
43 (79) & 50023-01-22-00 &  51695.340412 & $ 1.05 \pm 0.02$ & $ 1.98
\pm 0.13$ & $ 1.136 \pm 0.012$ & $ 2.00 \pm 0.06$ & $ 3.02 \pm
0.03$ \\ 
44 (80) & 50023-01-23-00 &  51697.479960 & $ 1.01 \pm 0.02$ & $ 2.10
\pm 0.14$ & $ 1.11 \pm 0.02$ & $ 2.02 \pm 0.12$ & $ 3.15 \pm
0.04$ \\ 
45 (82) & 50030-03-02-00 &  51942.946777 & $ 0.94 \pm 0.02$ & $ 1.49
\pm 0.11$ & $ 0.880 \pm 0.010$ & $ 2.04 \pm 0.06$ & $ 1.85 \pm
0.02$ \\ 
46 (83) & 50030-03-03-02 &  51949.126690 & $ 1.09 \pm 0.03$ & $ 2.3
\pm 0.2$ & $ 1.19 \pm 0.02$ & $ 1.87 \pm 0.08$ & $ 1.80 \pm
0.03$ \\ 
47 (85) & 50030-03-04-00 &  52007.613823 & $ 0.82 \pm 0.04$ & $ 1.06
\pm 0.13$ & $ 1.010 \pm 0.008$ & $ 2.10 \pm 0.04$ & $ 3.85 \pm
0.03$ \\ 
48 (86) & 50030-03-04-02 &  52008.087790 & $ 0.842 \pm 0.015$ & $ 1.08
\pm 0.06$ & $ 0.906 \pm 0.007$ & $ 2.04 \pm 0.04$ & $ 4.33 \pm
0.03$ \\ 
49 (87) & 50030-03-05-03 &  52024.438632 & $ 1.10 \pm 0.03$ & $ 2.22
\pm 0.15$ & $ 1.18 \pm 0.02$ & $ 1.97 \pm 0.08$ & $ 3.11 \pm
0.04$ \\ 
50 (88) & 50030-03-05-02 &  52024.696132 & $ 1.05 \pm 0.02$ & $ 2.8
\pm 0.2$ & $ 1.20 \pm 0.02$ & $ 1.92 \pm 0.08$ & $ 2.98 \pm
0.05$ \\ 
51 (89) & 60029-02-01-00 &  52056.408693 & $ 1.07 \pm 0.03$ & $ 1.82
\pm 0.15$ & $ 0.959 \pm 0.008$ & $ 2.03 \pm 0.04$ & $ 4.42 \pm
0.04$ \\ 
52 (91) & 50030-03-06-00 &  52112.254234 & $ 1.06 \pm 0.02$ & $ 2.05
\pm 0.12$ & $ 1.036 \pm 0.012$ & $ 1.97 \pm 0.05$ & $ 1.83 \pm
0.02$ \\
\hline\hline\\
\end{tabular}
\end{minipage}
\end{table*}

\begin{table*}
\begin{minipage}{170mm}
\contcaption{}
\label{bigtable4}
\begin{tabular}{lccccccc}
\hline\hline

B\# (G8\#) & OBSID & Date (MJD) & $\Ftd$ & $K_{\rm td/2} / K_{\rm td}$ & HC &
SC & $\Fper$\\
\hline
53 (92) & 50030-03-06-02 &  52112.585290 & $ 1.05 \pm 0.05$ & $ 1.9
\pm 0.2$ & $ 1.085 \pm 0.013$ & $ 1.96 \pm 0.06$ & $ 1.76 \pm
0.02$ \\ 
54 (97) & 70028-01-01-07 &  52336.211931 & $ 0.82 \pm 0.02$ & $ 1.24
\pm 0.08$ & $ 1.04 \pm 0.010$ & $ 2.07 \pm 0.05$ & $ 3.25 \pm
0.03$ \\ 
55 (100) & 70028-01-01-02 &  52337.098108 & $ 0.85 \pm 0.02$ & $ 1.39
\pm 0.11$ & $ 1.02 \pm 0.02$ & $ 2.02 \pm 0.08$ & $ 3.22 \pm
0.05$ \\ 
56 (101) & 70028-01-01-00 &  52337.946352 & $ 0.90 \pm 0.03$ & $ 1.52
\pm 0.13$ & $ 1.027 \pm 0.014$ & $ 2.06 \pm 0.10$ & $ 3.08 \pm
0.04$ \\ 
57 (102) & 70028-01-01-00 &  52338.091517 & $ 0.84 \pm 0.02$ & $ 1.19
\pm 0.09$ & $ 1.02 \pm 0.03$ & $ 1.8 \pm 0.3$ & $ 3.18 \pm
0.15$ \\ 
58 (104) & 70028-01-01-12 &  52338.416621 & $ 0.88 \pm 0.02$ & $ 1.60
\pm 0.11$ & $ 1.097 \pm 0.014$ & $ 2.04 \pm 0.09$ & $ 2.94 \pm
0.03$ \\ 
59  & 92023-03-03-00 &  53802.039063 & $ 0.90 \pm 0.03$ & $ 1.52 \pm
0.13$ & $ 1.108 \pm 0.012$ & $ 2.01 \pm 0.07$ & $ 3.43 \pm
0.03$ \\ 
60  & 92023-03-35-00 &  53866.028019 & $ 0.99 \pm 0.02$ & $ 1.86 \pm
0.12$ & $ 1.107 \pm 0.012$ & $ 2.00 \pm 0.05$ & $ 3.51 \pm
0.03$ \\ 
61  & 92023-03-02-10 &  53996.503670 & $ 0.95 \pm 0.02$ & $ 1.9 \pm
0.2$ & $ 1.17 \pm 0.03$ & $ 1.9 \pm 0.3$ & $ 3.9 \pm
0.2$ \\ 
62  & 92023-03-06-10 &  54004.550617 & $ 0.86 \pm 0.02$ & $ 1.10 \pm
0.10$ & $ 1.046 \pm 0.009$ & $ 2.10 \pm 0.05$ & $ 5.99 \pm
0.05$ \\ 
63  & 92023-03-16-10 &  54024.205142 & $ 0.83 \pm 0.02$ & $ 0.82 \pm
0.07$ & $ 0.957 \pm 0.015$ & $ 1.95 \pm 0.12$ & $ 3.49 \pm
0.07$ \\ 
64  & 92023-03-20-10 &  54032.060987 & $ 0.88 \pm 0.02$ & $ 0.86 \pm
0.05$ & $ 0.870 \pm 0.012$ & $ 1.96 \pm 0.11$ & $ 4.54 \pm
0.08$ \\ 
65  & 92023-03-27-10 &  54046.469298 & $ 0.77 \pm 0.02$ & $ 1.28 \pm
0.10$ & $ 0.961 \pm 0.011$ & $ 1.98 \pm 0.09$ & $ 5.93 \pm
0.07$ \\ 
66  & 92023-03-31-10 &  54054.205976 & $ 0.93 \pm 0.03$ & $ 1.51 \pm
0.12$ & $ 1.137 \pm 0.011$ & $ 2.02 \pm 0.06$ & $ 3.71 \pm
0.04$ \\ 
67  & 92023-03-34-10 &  54120.259567 & $ 1.04 \pm 0.07$ & $ 2.0 \pm
0.3$ & $ 1.131 \pm 0.011$ & $ 1.97 \pm 0.05$ & $ 2.51 \pm
0.02$ \\ 
68  & 92023-03-44-10 &  54140.744318 & $ 0.90 \pm 0.02$ & $ 1.50 \pm
0.10$ & $ 1.121 \pm 0.011$ & $ 2.03 \pm 0.05$ & $ 3.91 \pm
0.04$ \\ 
69  & 92023-03-44-00 &  54166.219564 & $ 0.98 \pm 0.03$ & $ 2.1 \pm
0.2$ & $ 1.147 \pm 0.014$ & $ 2.01 \pm 0.06$ & $ 3.59 \pm
0.04$ \\ 
70  & 92023-03-70-00 &  54226.783693 & $ 0.93 \pm 0.02$ & $ 2.24 \pm
0.15$ & $ 1.159 \pm 0.011$ & $ 2.05 \pm 0.06$ & $ 3.70 \pm
0.04$ \\ 
71  & 92023-03-71-00 &  54228.071109 & $ 0.90 \pm 0.02$ & $ 1.69 \pm
0.14$ & $ 1.097 \pm 0.012$ & $ 2.06 \pm 0.05$ & $ 4.31 \pm
0.04$ \\ 
72  & 92023-03-73-00 &  54230.488444 & $ 0.94 \pm 0.02$ & $ 1.55 \pm
0.12$ & $ 1.124 \pm 0.012$ & $ 2.04 \pm 0.06$ & $ 4.57 \pm
0.05$ \\ 
73  & 92023-03-66-10 &  54234.897703 & $ 0.90 \pm 0.03$ & $ 1.9 \pm
0.2$ & $ 1.15 \pm 0.03$ & $ 2.0 \pm 0.2$ & $ 5.05 \pm
0.11$ \\ 
74  & 92023-03-83-10 &  54268.409381 & $ 1.00 \pm 0.03$ & $ 1.9 \pm
0.2$ & $ 1.009 \pm 0.010$ & $ 2.05 \pm 0.06$ & $ 4.14 \pm
0.04$ \\ 
75  & 95337-01-02-00 &  55473.926815 & $ 0.96 \pm 0.02$ & $ 1.15 \pm
0.07$ & $ 0.785 \pm 0.011$ & $ 2.03 \pm 0.07$ & $ 2.38 \pm
0.04$ \\ 
76  & 96322-01-05-02 &  55840.957370 & $ 1.01 \pm 0.04$ & $ 1.6 \pm
0.2$ & $ 1.19 \pm 0.02$ & $ 2.00 \pm 0.08$ & $ 3.60 \pm
0.05$ \\ 
77  & 96322-01-05-00 &  55841.140924 & $ 0.99 \pm 0.02$ & $ 1.91 \pm
0.14$ & $ 1.20 \pm 0.02$ & $ 2.00 \pm 0.10$ & $ 3.62 \pm
0.05$ \\ 
78  & 96322-01-05-00 &  55841.301314 & $ 1.03 \pm 0.03$ & $ 2.2 \pm
0.2$ & $ 1.18 \pm 0.06$ & $ 1.7 \pm 0.4$ & $ 4.0 \pm
0.3$ \\ 
79  & 96322-01-05-00 &  55841.477828 & $ 1.04 \pm 0.03$ & $ 1.72 \pm
0.13$ & $ 1.20 \pm 0.03$ & $ 1.8 \pm 0.2$ & $ 3.73 \pm
0.12$ \\
\hline
\multicolumn{8}{c}{4U 1735--44} \\
\hline
 1 (3) & 20084-01-02-04 &  50693.542191 & $ 0.390 \pm 0.008$ & $ 0.90 \pm
0.07$ & $ 0.84 \pm 0.02$ & $ 1.7 \pm 0.2$ & $ 5.2 \pm
0.2$ \\ 
 2 (6) & 30056-02-01-00 &  50963.430507 & $ 0.354 \pm 0.009$ & $ 1.02 \pm
0.10$ & $ 0.949 \pm 0.010$ & $ 1.88 \pm 0.05$ & $ 3.19 \pm
0.03$ \\ 
 3 (7) & 30056-02-01-00 &  50963.490137 & $ 0.334 \pm 0.007$ & $ 0.99 \pm
0.08$ & $ 0.938 \pm 0.009$ & $ 1.84 \pm 0.06$ & $ 3.24 \pm
0.04$ \\ 
 4 (8) & 30056-02-01-00 &  50963.547509 & $ 0.283 \pm 0.006$ & $ 0.92 \pm
0.08$ & $ 0.933 \pm 0.007$ & $ 1.86 \pm 0.03$ & $ 3.20 \pm
0.02$ \\ 
 5 (10) & 40030-02-01-00 &  51347.126951 & $ 0.340 \pm 0.010$ & $ 1.38
\pm 0.12$ & $ 0.94 \pm 0.04$ & $ 1.9 \pm 0.5$ & $ 3.6 \pm
0.4$ \\ 
 6 (11) & 40031-02-01-04 &  51348.108487 & $ 0.368 \pm 0.009$ & $ 0.88
\pm 0.08$ & $ 0.87 \pm 0.03$ & $ 1.5 \pm 0.3$ & $ 4.4 \pm
0.4$ \\ 
 7  & 91025-01-10-01 &  54727.956277 & $ 0.360 \pm 0.012$ & $ 0.95 \pm
0.12$ & $ 0.926 \pm 0.013$ & $ 1.85 \pm 0.08$ & $ 3.44 \pm
0.06$ \\ 
 8  & 91025-01-11-00 &  54728.905009 & $ 0.393 \pm 0.012$ & $ 0.85 \pm
0.12$ & $ 0.901 \pm 0.010$ & $ 1.93 \pm 0.06$ & $ 4.30 \pm
0.05$ \\ 
 9  & 93200-01-01-03 &  54879.335374 & $ 0.438 \pm 0.014$ & $ 1.00 \pm
0.12$ & $ 1.06 \pm 0.03$ & $ 2.04 \pm 0.14$ & $ 2.10 \pm
0.04$ \\ 
10  & 93200-01-01-02 &  54879.855383 & $ 0.339 \pm 0.013$ & $ 1.8 \pm
0.2$ & $ 1.068 \pm 0.012$ & $ 1.93 \pm 0.03$ & $ 2.37 \pm
0.02$ \\ 
11  & 93200-01-02-00 &  54974.580901 & $ 0.407 \pm 0.013$ & $ 1.16 \pm
0.14$ & $ 0.894 \pm 0.014$ & $ 1.94 \pm 0.11$ & $ 3.59 \pm
0.07$ \\ 
12  & 93200-01-02-00 &  54974.783684 & $ 0.377 \pm 0.012$ & $ 0.83 \pm
0.11$ & $ 0.87 \pm 0.02$ & $ 2.0 \pm 0.2$ & $ 3.8 \pm
0.2$ \\ 
13  & 93200-01-03-01 &  55123.323892 & $ 0.446 \pm 0.013$ & $ 0.75 \pm
0.10$ & $ 0.95 \pm 0.02$ & $ 1.86 \pm 0.11$ & $ 3.08 \pm
0.07$ \\ 
\hline
\multicolumn{8}{c}{4U 1820--30} \\
\hline
 1 (1) & 20075-01-05-00 &  50570.731795 & $ 0.59 \pm 0.02$ & $ 1.55 \pm
0.14$ & $ 1.033 \pm 0.007$ & $ 1.79 \pm 0.02$ & $ 3.683 \pm
0.015$ \\ 
 2 (2) & 40017-01-24-00 &  52794.738826 & $ 0.621 \pm 0.015$ & $ 1.53 \pm
0.12$ & $ 1.161 \pm 0.009$ & $ 1.97 \pm 0.03$ & $ 2.93 \pm
0.02$ \\ 
 3 (3) & 70030-03-04-01 &  52802.076265 & $ 0.59 \pm 0.02$ & $ 1.38 \pm
0.12$ & $ 1.145 \pm 0.009$ & $ 1.97 \pm 0.03$ & $ 2.92 \pm
0.02$ \\ 
 4 (4) & 70030-03-05-01 &  52805.896358 & $ 0.609 \pm 0.014$ & $ 1.45 \pm
0.13$ & $ 1.145 \pm 0.010$ & $ 1.95 \pm 0.02$ & $ 3.81 \pm
0.02$ \\ 
 5 (5) & 90027-01-03-05 &  53277.439257 & $ 0.628 \pm 0.013$ & $ 1.46 \pm
0.11$ & $ 1.015 \pm 0.008$ & $ 1.81 \pm 0.04$ & $ 3.83 \pm
0.04$ \\ 
 6  & 94090-01-01-02 &  54948.821939 & $ 0.60 \pm 0.02$ & $ 1.6 \pm
0.2$ & $ 1.111 \pm 0.015$ & $ 1.97 \pm 0.04$ & $ 3.58 \pm
0.04$ \\ 
 7  & 94090-01-01-05 &  54950.703513 & $ 0.58 \pm 0.02$ & $ 1.36 \pm
0.13$ & $ 1.18 \pm 0.02$ & $ 1.96 \pm 0.11$ & $ 3.44 \pm
0.05$ \\ 
 8  & 94090-01-02-03 &  54956.775426 & $ 0.59 \pm 0.02$ & $ 1.43 \pm
0.15$ & $ 1.24 \pm 0.02$ & $ 2.01 \pm 0.11$ & $ 3.28 \pm
0.05$ \\ 
 9  & 94090-01-02-02 &  54958.740672 & $ 0.61 \pm 0.02$ & $ 1.50 \pm
0.15$ & $ 1.25 \pm 0.03$ & $ 2.1 \pm 0.2$ & $ 3.09 \pm
0.09$ \\ 
10  & 94090-01-04-00 &  54978.322182 & $ 0.63 \pm 0.02$ & $ 1.8 \pm
0.2$ & $ 1.27 \pm 0.03$ & $ 1.9 \pm 0.2$ & $ 3.75 \pm
0.10$ \\ 
11  & 94090-01-04-01 &  54978.495588 & $ 0.61 \pm 0.02$ & $ 1.28 \pm
0.14$ & $ 1.33 \pm 0.02$ & $ 2.03 \pm 0.06$ & $ 3.52 \pm
0.03$ \\ 
12  & 94090-01-05-00 &  54981.187938 & $ 0.614 \pm 0.015$ & $ 1.57 \pm
0.12$ & $ 1.28 \pm 0.02$ & $ 2.00 \pm 0.12$ & $ 4.05 \pm
0.06$ \\ 
13  & 94090-02-01-00 &  54994.534879 & $ 0.58 \pm 0.02 $ & $ 1.15 \pm
0.13$ & $ 1.185 \pm 0.012$ & $ 1.99 \pm 0.03$ & $ 5.84 \pm
0.06$ \\ 
14  & 94090-02-01-00 &  54994.613713 & $ 0.53 \pm 0.02$ & $ 1.03 \pm
0.11$ & $ 1.130 \pm 0.015$ & $ 1.91 \pm 0.09$ & $ 6.10 \pm
0.12$ \\ 
15  & 96090-01-01-00 &  55624.881378 & $ 0.58 \pm 0.02$ & $ 1.41 \pm
0.14$ & $ 1.15 \pm 0.02$ & $ 2.01 \pm 0.11$ & $ 2.95 \pm
0.05$ \\ 
16  & 96090-01-01-02 &  55626.774306 & $ 0.55 \pm 0.02$ & $ 1.28 \pm
0.14$ & $ 0.912 \pm 0.009$ & $ 1.78 \pm 0.07$ & $ 4.67 \pm
0.08$ \\
\hline\hline\\
\end{tabular}
\end{minipage}
\end{table*}

\begin{table*}
\begin{minipage}{170mm}
\contcaption{}
\label{bigtable5}
\begin{tabular}{lccccccc}
\hline\hline

B\# (G8\#) & OBSID & Date (MJD) & $\Ftd$ & $K_{\rm td/2} / K_{\rm td}$ & HC &
SC & $\Fper$\\
\hline
\multicolumn{8}{c}{Aql X--1} \\
\hline
 1 (4) & 20098-03-08-00 &  50508.977504 & $ 1.22 \pm 0.02$ & $ 1.32 \pm
0.07$ & $ 0.505 \pm 0.008$ & $ 1.44 \pm 0.02$ & $ 1.37 \pm
0.02$ \\ 
 2 (5) & 20092-01-05-00 &  50696.524280 & $ 1.21 \pm 0.02$ & $ 0.95 \pm
0.05$ & $ 0.521 \pm 0.007$ & $ 1.43 \pm 0.06$ & $ 3.56 \pm
0.07$ \\ 
 3 (6) & 20092-01-05-03 &  50699.400018 & $ 0.650 \pm 0.009$ & $ 1.17 \pm
0.07$ & $ 0.503 \pm 0.005$ & $ 1.53 \pm 0.02$ & $ 3.31 \pm
0.02$ \\ 
 4 (10) & 40047-03-02-00 &  51332.780597 & $ 1.31 \pm 0.02$ & $ 1.12
\pm 0.06$ & $ 0.562 \pm 0.012$ & $ 1.44 \pm 0.08$ & $ 2.60 \pm
0.08$ \\ 
 5 (11) & 40047-03-06-00 &  51336.591439 & $ 1.20 \pm 0.04$ & $ 1.9
\pm 0.2$ & $ 0.467 \pm 0.012$ & $ 1.54 \pm 0.07$ & $ 1.05 \pm
0.03$ \\ 
 6 (19) & 50049-02-13-01 &  51856.157590 & $ 0.75 \pm 0.02$ & $ 1.63
\pm 0.13$ & $ 0.526 \pm 0.006$ & $ 1.59 \pm 0.06$ & $ 7.77 \pm
0.15$ \\ 
 7 (28) & 60429-01-06-00 &  52324.991248 & $ 1.22 \pm 0.02$ & $ 0.93
\pm 0.05$ & $ 0.553 \pm 0.006$ & $ 1.52 \pm 0.03$ & $ 2.53 \pm
0.03$ \\ 
 8 (29) & 70069-03-02-03 &  52347.182990 & $ 0.668 \pm 0.013$ & $ 1.9
\pm 0.2$ & $ 0.514 \pm 0.006$ & $ 1.55 \pm 0.05$ & $ 3.71 \pm
0.06$ \\ 
 9  & 92438-01-02-01 &  54259.248574 & $ 1.18 \pm 0.03$ & $ 3.7 \pm
0.3$ & $ 1.23 \pm 0.04$ & $ 1.67 \pm 0.05$ & $ 0.355 \pm
0.006$ \\ 
10  & 93405-01-03-07 &  54365.807784 & $ 0.85 \pm 0.03$ & $ 0.77 \pm
0.09$ & $ 0.583 \pm 0.009$ & $ 1.42 \pm 0.11$ & $ 6.7 \pm
0.3$ \\ 
11  & 94076-01-05-02 &  55157.140116 & $ 1.11 \pm 0.02$ & $ 1.12 \pm
0.07$ & $ 0.538\pm 0.006$ & $ 1.53 \pm 0.05$ & $ 5.53 \pm
0.13$ \\ 
12  & 96440-01-09-07 &  55904.228389 & $ 1.13 \pm 0.03$ & $ 2.0 \pm
0.2$ & $ 0.56 \pm 0.02$ & $ 1.44 \pm 0.08$ & $ 1.26 \pm
0.04$ \\
\hline
\multicolumn{8}{c}{HETE J1900.1--2455} \\
\hline
 1 (1) & 91059-03-01-04 &  53572.959405 & $ 1.234 \pm 0.014$ & $ 4.5 \pm
0.2$ & $ 1.6 \pm 0.2$ & $ 1.59 \pm 0.08$ & $ 0.723 \pm
0.011$ \\ 
 2 (2) & 92049-01-07-00 &  53814.482712 & $ 1.08 \pm 0.02$ & $ 3.5 \pm
0.2$ & $ 1.47 \pm 0.02$ & $ 1.63 \pm 0.14$ & $ 1.26 \pm
0.03$ \\ 
 3  & 93030-01-23-00 &  54439.248861 & $ 1.17 \pm 0.03$ & $ 3.6 \pm
0.3$ & $ 1.52 \pm 0.03$ & $ 1.71 \pm 0.08$ & $ 1.18 \pm
0.02$ \\ 
 4  & 93030-01-25-00 &  54506.856847 & $ 1.12 \pm 0.03$ & $ 3.3 \pm
0.2$ & $ 1.38 \pm 0.04$ & $ 1.7 \pm 0.2$ & $ 1.48 \pm
0.06$ \\ 
 5  & 94030-01-09-00 &  54923.374237 & $ 0.98 \pm 0.02$ & $ 1.05 \pm
0.08$ & $ 0.668 \pm 0.011$ & $ 1.75 \pm 0.03$ & $ 1.754 \pm
0.010$ \\ 
 6  & 94028-01-01-03 &  54925.797119 & $ 1.27 \pm 0.03$ & $ 2.3 \pm
0.2$ & $ 1.03 \pm 0.07$ & $ 1.71 \pm 0.05$ & $ 1.11 \pm
0.02$ \\ 
 7  & 94030-01-41-00 &  55145.515365 & $ 1.14 \pm 0.03$ & $ 2.7 \pm
0.2$ & $ 1.28 \pm 0.03$ & $ 1.72 \pm 0.11$ & $ 1.30 \pm
0.03$ \\ 
 8  & 95030-01-23-00 &  55384.878919 & $ 1.18 \pm 0.03$ & $ 3.4 \pm
0.2$ & $ 1.52 \pm 0.03$ & $ 1.70 \pm 0.05$ & $ 0.930 \pm
0.008$ \\ 
 9  & 95030-01-34-00 &  55459.229332 & $ 1.06 \pm 0.03$ & $ 3.72 \pm
0.3$ & $ 1.45 \pm 0.02$ & $ 1.73 \pm 0.03$ & $ 1.323 \pm
0.012$ \\ 
10  & 96030-01-35-00 &  55833.989462 & $ 1.16 \pm 0.03$ & $ 3.3 \pm
0.2$ & $ 1.36 \pm 0.02$ & $ 1.77 \pm 0.07$ & $ 1.31 \pm
0.02$ \\
\hline
\multicolumn{8}{c}{SAX J1808.4--3658} \\
\hline
 1  & 93027-01-01-08 &  54732.708829 & $ 2.23 \pm 0.05$ & $ 2.4 \pm
0.2$ & $ 1.32 \pm 0.02$ & $ 1.70 \pm 0.03$ & $ 1.82 \pm
0.02$ \\ 
 2  & 93027-01-01-07 &  54733.844037 & $ 2.27 \pm 0.05$ & $ 2.4 \pm
0.2$ & $ 1.32 \pm 0.02$ & $ 1.64 \pm 0.06$ & $ 2.02 \pm
0.02$ \\ 
 3  & 96027-01-01-07 &  55873.917044 & $ 2.68 \pm 0.05$ & $ 2.04 \pm
0.13$ & $ 1.22 \pm 0.02$ & $ 1.52 \pm 0.04$ & $ 2.38 \pm
0.03$ \\
\hline\hline\\
\end{tabular}
\end{minipage}
\end{table*}

\end{document}